\PassOptionsToPackage{table}{xcolor}

\documentclass{article}

\usepackage{microtype}
\usepackage{graphicx}
\usepackage{booktabs} 
\usepackage{enumitem}
\usepackage{listings}
\usepackage{svg}
\svgsetup{inkscapearea=page}
\usepackage{mathtools}
\usepackage{multirow}
\usepackage{hyperref}
\usepackage{pifont}
\usepackage{subcaption}
\usepackage{makecell}
\usepackage{titlesec}
\titlespacing*{\section}{0pt}{2ex}{0.5ex}
\titlespacing*{\subsection}{0pt}{1.2ex}{0.5ex}

\newcommand{\proj}{\textsc{JigsawServe}}

\def\checkmark{\includegraphics[width=1.5em]{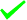}}

\def\cross{\includegraphics[width=0.75em]{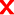}}

\usepackage[nohyperref, accepted]{mlsys2025}

\begin{document}

\twocolumn[

\mlsystitle{Serving Compound Inference Systems on Datacenter GPUs}



\mlsyssetsymbol{equal}{*}

\begin{mlsysauthorlist}
\mlsysauthor{Sriram Devata}{uiuc}
\mlsysauthor{Rahul Singh}{uiuc}
\mlsysauthor{Sarita Adve}{uiuc}
\end{mlsysauthorlist}

\mlsysaffiliation{uiuc}{Siebel School of Computing and Data Science, University of Illinois Urbana-Champaign, Champaigh, IL, USA}

\mlsyscorrespondingauthor{Sriram Devata}{sdevata2@illinois.edu}

\mlsyskeywords{Machine Learning, MLSys}

\vskip 0.3in

\begin{abstract}

    Applications in emerging domains such as XR are being built as compound inference systems, where multiple ML models are composed in the form of a task graph to service each request. 
    Serving these compound systems efficiently raises two questions: how to apportion end-to-end latency and accuracy budgets between different tasks in a compound inference system, and how to allocate resources effectively for different models with varying resource requirements.
    We present \proj{}, the first serving framework that jointly optimizes for latency, accuracy, and cost in terms of GPU resources by adaptively choosing model variants and performing fine-grained resource allocation by spatially partitioning the GPUs for each task of a compound inference system.
    Analytical evaluation of a system with a large number of GPUs shows that \proj{} can increase the maximum serviceable demand (in requests per second) by $11.3\times$ when compared to the closest prior work. Our empirical evaluation shows that for a large range of scenarios, \proj{} consumes only 43.3\% of the available GPU resources while meeting accuracy SLOs with less than 0.6\% latency SLO violations. All of the features in \proj{} contribute to this high efficiency -- sacrificing any one feature of accuracy scaling, GPU spatial partitioning, or task-graph-informed resource budgeting significantly reduces efficiency.
\end{abstract}

]



{\let\thefootnote\relax\footnotetext{\textsuperscript{1}Siebel School of Computing and Data Science, University of Illinois Urbana-Champaign \\
Email: \{sdevata2, rahuls10, sadve\}@illinois.edu
}
}

\section{Introduction} \label{introduction}

As machine learning (ML) inference becomes a dominant workload for datacenter resources~\cite{inferencecostnvidia,energydatacentergoogle,metacarbonfootprint}, there is growing emphasis on maximizing the efficiency of inference serving systems.
Compound inference systems \cite{compoundaisystems} are becoming increasingly important in many emerging domains such as multi-agent systems \cite{multiagentsystem} and extended reality (XR) \cite{xrbench}. Such systems compose multiple ML models, each performing a specific task, to provide complex functionalities, and present new challenges and opportunities for increased efficiency. This work focuses on efficient serving of compound inference systems on datacenter GPUs, where each request requires invoking a directed acyclic graph (DAG) of ML inference tasks, which are potentially dynamically determined.

We focus on two aspects that are unique to compound inference systems. First, we expect that any latency and accuracy 
Service Level Objectives (SLOs) are provided for the end-to-end compound inference system and need to be apportioned among the individual tasks. This affords new sources of flexibility for the serving system to choose among multiple model variants (with different latencies and accuracies) and GPU resource allocation for each task, when compared to a single task system with a given SLO constraint. Second, the ML model variants across the various tasks in a compound inference system have different resource requirements, motivating finer-grained heterogeneity in the optimal resource choice for the different tasks. Recent GPU hardware provide spatial partitioning mechanisms that enable fine-grained allocation of GPU resources. For example, NVIDIA GPUs provide spatial partitioning mechanisms such as Multi-Process Service (MPS) \cite{nvidiamps} and Multi-Instance GPU (MIG) \cite{nvidiamig}.

We present \proj{} \footnote{The name highlights the features of \proj{} to choose model variants and use multiple small spatial partitions on GPUs, similar to how one can choose multiple small pieces that fit together in a jigsaw puzzle.}, the first work to jointly optimize for latency, accuracy, and fine-grained cost (in terms of GPU resources required) by adaptively choosing model variants and allocating fine-grained spatial partitions of GPUs for each task of a compound inference system, while preserving end-to-end accuracy and latency SLOs.

There is a plethora of literature on increasing the efficiency of serving systems for requests with a single model inference, including systems that perform accuracy scaling and spatial partitioning \cite{nexus, infaas, migserving, gpulet, proteus, parvagpu}. Compound inference systems have received relatively less attention. Table \ref{tab:featuresrelatedworks} summarizes key prior works that consider resource budgeting for compound inference systems, accuracy scaling through different model variants of a task, and/or spatial partitioning of GPU resources. The darker grayed systems do not consider compound inferences but are shown for completeness. The lighter grayed systems do consider compound inferences but in a limited way -- they apportion latency across different tasks (only pipelines for IPA \cite{ipa}), but do not apportion accuracy or consider accuracy scaling or GPU partitioning for higher efficiency. Of the remaining systems that consider compound inferences, DREAM \cite{dream} is a single user, multi-accelerator system and IPA does not consider GPUs, both presenting different use cases from our datacenter GPU efficiency scenario; consequently, neither considers the fine-grained spatial partitioning opportunity from GPUs. Loki \cite{loki} is the closest to our goals, but does not consider GPU partitioning and considers accuracy scaling only if GPU resources are exhausted. Our evaluations show that GPU partitioning presents the most significant efficiency opportunity and jointly optimizing for GPU resources and accuracy for compound inference systems is critical to highest efficiency. 


Specifically, we make the following contributions.

\noindent
\begin{enumerate}[topsep=1pt,noitemsep, leftmargin=1em]
    \item We introduce \proj{}
    , a novel framework to jointly optimize latency, accuracy, and fine-grained cost for compound inference systems on datacenter GPUs. \proj{} uniquely combines: (1) per-task accuracy scaling via choosing model variants, (2) GPU spatial partitioning, and (3) task-graph-informed latency and accuracy SLO budgeting.
    
    \item Our analytical evaluation compares the various combinations of the features in Table \ref{tab:featuresrelatedworks}. We find that spatial partitioning (S) delivers the highest standalone gain in serving capacity for the same GPU resources (5.25$\times$ over Unopt), compared to accuracy scaling (A: 1.6$\times$) and task-graph-informed budgeting (T: 1.1$\times$). \proj{}'s full integration of the three features (A+S+T) achieves 21.6$\times$ capacity, surpassing combinations of S with A or T in S+A (12.1$\times$) and S+T (7.8$\times$).
    
    \item We emphasize that no prior work has evaluated GPU spatial partitioning for compound inference systems. Systems without S (A, T, A+T) show significantly lower serving capacity than any system with S. A+T, in particular, is closest to (and explores a larger search space than) prior work of Loki \cite{loki}. We find that \proj{} allows 11.3$\times$ higher capacity than A+T.
    
    \item Our empirical evaluation compares the four top performing systems (S+T, A+T, S+A, and \proj{}) under a wide range of compound inference systems and workload demand conditions. \proj{} shows the best overall performance, using just 43.3\% of the available GPU resources on average, respecting the accuracy SLO, and showing less than 0.6\% average SLO violations. Alternatives suffer significantly: S+T/A+T exceed 10\% violations and have 2$\times$ resource usage in at least one case; S+A (which does not exist in prior work) requires 33\% more resources than \proj{} with 6.7\% SLO violations.
\end{enumerate}

Overall, our work proposes a novel serving framework for compound inference systems, performs a detailed comparative evaluation of efficiency-enhancing features, and shows that the combination of all features we explore provides significant efficiency improvement relative to any subset of features.

\begin{table}[t]
\caption{Summary of key related work.
}
\label{tab:featuresrelatedworks}
\resizebox{0.49\textwidth}{!}{%
\begin{tabular}{|c|c|c|c|}
\hline
Related Work & \shortstack[c]{\textbf{Model variants for} \\ \textbf{accuracy scaling}} & \shortstack[c]{\textbf{Interference-free} \\ \textbf{spatial partitioning}} & \shortstack[c]{\textbf{Task-graph-informed} \\ \textbf{latency \& accuracy budgets}} \\
\hline
\rowcolor{gray} \shortstack[c]{INFaaS\\\cite{infaas}}           & \checkmark     & \cross       & \cross         \\
\rowcolor{gray} \shortstack[c]{MIG-serving\\\cite{migserving}}  & \cross         & \checkmark   & \cross         \\
\rowcolor{gray} \shortstack[c]{ParvaGPU\\\cite{parvagpu}}       & \cross         & \checkmark   & \cross         \\
\rowcolor{gray} \shortstack[c]{Clover\\\cite{clover}}           & \checkmark     & \checkmark   & \cross         \\
\rowcolor{lightgray} \shortstack[c]{Nexus\\\cite{nexus}}             & \cross         & \cross       & \Gape[0pt][2pt]{\makecell{Only apportions latency}}     \\
\rowcolor{lightgray} \shortstack[c]{GPUlet\\\cite{gpulet}}           & \cross         & \cross       & \Gape[0pt][2pt]{\makecell{Only apportions latency}}     \\
\rowcolor{lightgray} \shortstack[c]{DREAM\\\cite{dream}}             & \checkmark     & \cross       & \Gape[0pt][2pt]{\makecell{Considers a single-client\\multi-accelerator system}}     \\
\rowcolor{lightgray} \shortstack[c]{IPA\\\cite{ipa}}                 & \checkmark     & \cross       & \Gape[0pt][2pt]{\shortstack[c]{Only considers pipelines. \\ Apportions CPU cores.}}     \\
\shortstack[c]{Loki\\\cite{loki}}               & \checkmark     & \cross       & \checkmark     \\
\textbf{\proj{}} & \checkmark & \checkmark & \checkmark \\
\hline
\end{tabular}
}
\end{table}




\section{Background} \label{background}


\textbf{Compound inference systems.} ML-based applications are evolving from relying on the inference of a single monolithic ML model to compound inference, where multiple ML models each performing a particular task are composed to provide complex functionalities \cite{xrbench}.
Similar to single ML model inference, the demand (in requests per second) for compound inference systems can vary over time. The serving system must increase the allocated GPU resources when demand rises and reduce them when demand falls, avoiding both SLO violations and over‑provisioning.

Tasks in a compound inference system often have unequal throughput demands—for instance, an object detector may output several detections per image, each triggering additional downstream inferences. Prior works \cite{slopt, loki} have noted this multiplicative behavior between tasks and used it to divide latency budgets, but none have jointly optimized model variants, resource allocation, and accuracy across the entire system. A serving system for compound inference must incorporate per-task multiplicative factors to estimate each task’s throughput and use it for task-graph-informed latency, accuracy and resource budgeting. The system must ensure that every task receives sufficient resources while the overall system satisfies the latency, accuracy, and resource constraints

\textbf{Model variants.} 
Each task in a compound inference system can be executed by multiple model variants that perform the same function but differ in parameters, accuracy, latency, and resource cost. These variants may come from model families, compiler optimizations such as TVM or TensorRT, or techniques like quantization, pruning, and layer fusion. While allocating accuracy budget for each task, a serving system must meet the end-to-end accuracy threshold of the compound inference system. They key challenge is to identify a set of task variants that have minimal impact on end-to-end quality, while maximizing the throughput or cost benefit.

\textbf{GPU spatial partitioning mechanisms.} 
Recent generations of GPUs in GPU-based datacenters are large and are typically underutilized by most ML workloads. To address the underutilization, modern GPUs allow multiple workloads to be co-located on the same physical GPU through temporal and spatial sharing mechanisms such as MPS and MIG for NVIDIA GPUs. 
Workloads co-located with MPS share internal GPU memory resources, which can result in performance degradation through destructive interference between co-located workloads \cite{misompsformig}. MIG can divide a GPU into isolated \textit{MIG instances}, each formed by multiple GPU slices that contain a dedicated proportion of SMs, caches, GPU memory, and memory bandwidth. This hardware-supported isolation minimizes interference between workloads on different MIG instances.

To achieve even higher GPU utilization, we leverage NVIDIA GPUs' capability to use MPS within MIG instances \cite{migserving, miger, parvagpu}. We use the \textit{GPU segment} \cite{parvagpu} abstraction to describe these MPS-enabled MIG instances. Each GPU segment is described by its MIG instance size and the number of concurrent identical MPS processes running within it. Crucially, when we co-locate multiple GPU segments on one physical GPU, workloads in different GPU segments minimally interfere with each other due to MIG's hardware isolation.

\textbf{Configuration search space.} We define a \textit{configuration} as the combination of model variants, GPU segment types, and maximum batch sizes that the serving system selects for each model instance across all tasks. As Table \ref{tab:featuresrelatedworks} shows, \proj{} has all three features of performing accuracy scaling, enabling spatial partitioning, and performing task-graph-informed resource budgeting. For a compound inference system with $T$ tasks, $M$ model variants per task, $N$ instances for each task, $S$ GPU segment types, and $B$ possible batch sizes, the \textit{configuration search space} has $(M*S*B)^{N*T}$ possible configurations. When a serving system performs accuracy scaling and enables spatial partitioning while being task-graph-uninformed, the configuration search space has $T*(M*S*B)^N$ configurations due to per-task resource budgets. Without accuracy scaling ($M=1$) or spatial partitioning ($S=1$), the configuration search space reduces further.

\section{Design of \proj{}} \label{systemdesign}

\begin{figure}[tbp]
  \includegraphics[width=0.49\textwidth,page=1]{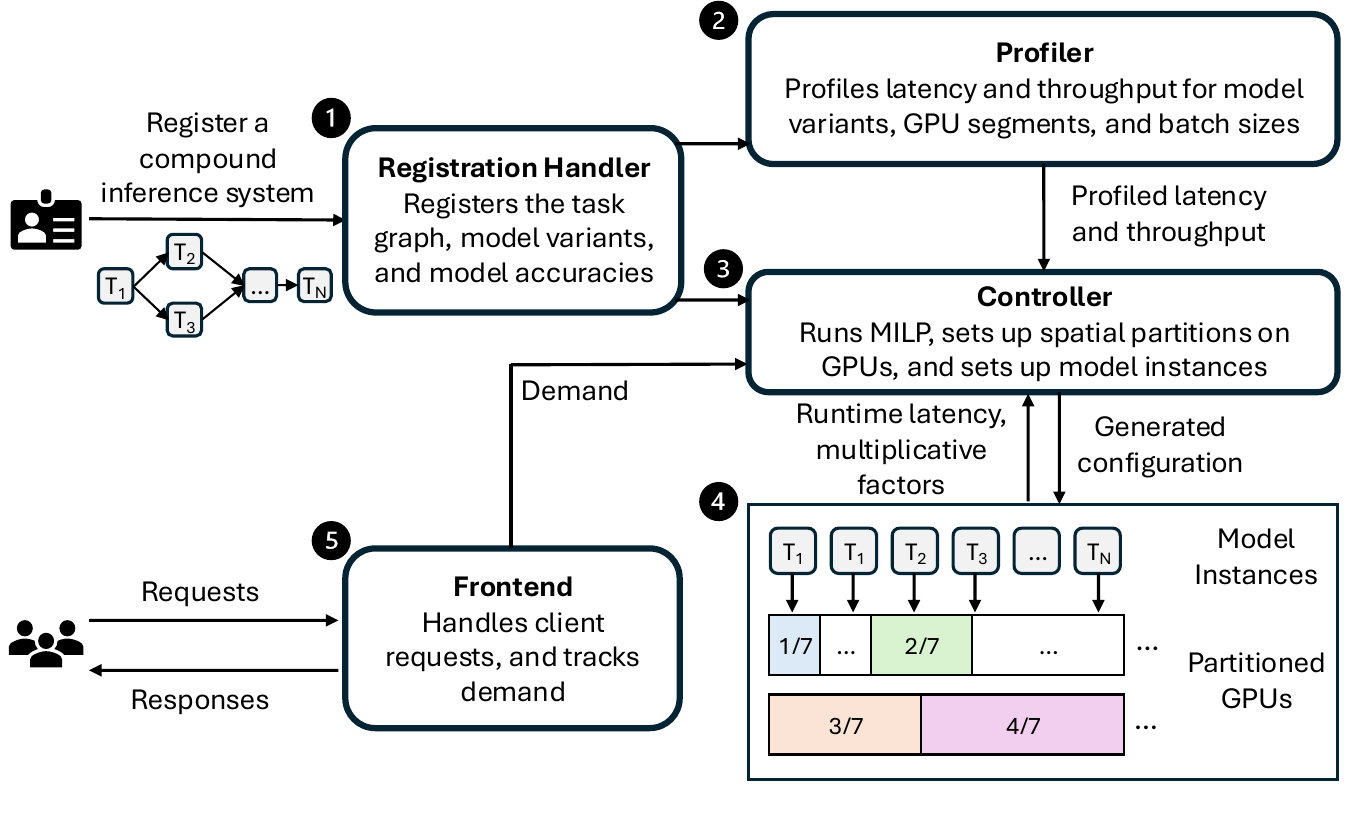}
  \vspace{-0.20in}
  \caption[System design of \proj{}.]{An overview of \proj{}. 
  }
  \label{fig:systemdesign}
    \vspace{-0.20in}
\end{figure}




\proj{} is made of four major components as shown in Figure \ref{fig:systemdesign}: a registration handler to register the compound inference system, a profiler that performs offline-profiling of each task, a controller to find the best configuration and set up model instances, and a frontend that clients can use to submit requests to the compound inference system. \S \ref{components} describes the components of \proj{} in detail. \S \ref{milp} describes the Mixed Integer Linear Programming Problem (MILP) formulation used by the controller. \S \ref{batchingearlydropping} details the batching and early drop mechanisms employed by each model instance in the compound inference system. 




\subsection{Components of \proj{}} \label{components}

\textbf{Registration handler}. Registering the compound inference system involves listing the tasks in the system, providing the weights of all model variants for each task along with their accuracies, describing the dependencies (edges) in the task graph, and specifying the end-to-end latency and accuracy SLO. The accuracy SLO is specified in terms of the acceptable drop in accuracy relative to the maximum achievable with the most accurate model variants. 

\textbf{Profiler}. After a compound inference system is registered with \proj{}, the profiler profiles the tasks in the system to find the 95th percentile latency and throughput (inferences per second) for all combinations of model variants, batch size, and GPU segment types. \proj{} considers the GPU segments formed by all possible MIG instance sizes with up to 4 concurrent MPS processes running the same model in each MIG instance to avoid exhaustive search and GPU out-of-memory issues \cite{miger, parvagpu}. The one-time profiling cost is compensated by sustained performance gains made possible by the profiling information. \proj{} also refines this data using actual runtime latency and throughput during the compound inference system execution.

\textbf{Controller}. The controller uses the profiled information to generate an optimal configuration to serve the required demand while satisfying latency and accuracy SLOs. We formulate the problem of finding the optimal configuration as an MILP, and use a commercial solver to find solutions for the MILP. We describe the MILP formulation in more detail in \S \ref{milp}.

The controller uses the generated configuration to set up spatial partitions on the GPUs and assigns the spatial partitions to the model instances. The controller uses a greedy rule-based bin-packing algorithm \cite{ibmmigpacking} to efficiently set up the MIG instances on the minimal number of GPUs. 
The controller then spawns workers for each model instance, assigns the MIG instances to the workers, and sets up inter-task data dependencies.

\textbf{Frontend}. The frontend collects requests from the clients, sets up the metadata for each request (request ID, and a deadline derived from the latency SLO), and routes the request to the first task of the compound inference system. The frontend also returns the output of the compound inference system to the client. The frontend continuously tracks the incoming demand and the SLO violation rate, triggering the controller to find a new configuration when the demand changes.

\subsection{MILP Formulation} \label{milp}

The controller formulates the problem of finding an optimal configuration as an MILP. Table \ref{tab:milpterms} summarizes the inputs and terms used for the MILP formulation. The inputs and profiling information are provided to the MILP. The intermediate variables are calculated within the MILP while evaluating candidate configurations. The decision variable candidate by the MILP is $M(t,v,s,b)$, which denotes for each task $t$, the number of instances of model variant $v$, running on GPU segment type $s$ with a maximum batch size $b$. $N(t,v,s,b)$ is a binary variable that denotes if there is at least 1 instance as described by the tuple $(t,v,s,b)$.
\begin{table}[h!tb]

\caption{Terms in the MILP.}
\label{tab:milpterms}
\resizebox{0.485\textwidth}{!}{%
\begin{tabular}{|cp{0.4\textwidth}|}
\hline
\textbf{Term} & \textbf{Description} \\
\hline
\hline
\multicolumn{2}{|c|}{\textbf{Inputs}} \\
\hline
$SLO_l$     & end-to-end latency SLO of the entire system \\
$SLO_a$     & threshold for the overall system accuracy, expressed as a fraction of the maximum possible accuracy \\
$A_{max}$ & the maximum overall system accuracy possible for an application \\
$R$         & demand (req/s) for the first task \\
$T$         & set of tasks in the compound inference system \\
$V_t$       & set of all model variants that perform task $t \in T$ \\
$S$         & set of all GPU segment types (MPS-enabled MIG instances) \\
$s_n$       & cost of GPU segment type $s$ in terms of GPU slices \\
$S_{avail}$ & number of total available GPU slices \\
$P$         & set of all paths that a request can take in the compound inference system inferred from the task graph \\
$P(t)$      & preceding tasks of task $t$ inferred from the task graph \\
$f_p$       & fraction of requests that take path $p$ in the compound inference system \\
$B$         & set of all possible batch sizes $[1,2,4,8,16,32,64,128]$ \\
\hline
\multicolumn{2}{|c|}{\textbf{Profiled Information}} \\
\hline
$L(t,v,s,b)$        & offline-profiled latency of model variant $v$ on GPU segment $s$ with batch size $b$ for task $t$ \\
$H(t,v,s,b)$        & offline-profiled throughput (req/s) of model variant $v$ on GPU segment $s$ with batch size $b$ for task $t$ \\
$F(t, v, t')$       & multiplicative factor of model variant $v$ in task $t$ to task $t'$ \\
$A(t, v)$           & accuracy of model variant $v$ for task $t$ \\
\hline
\multicolumn{2}{|c|}{\textbf{Intermediate Variables}} \\
\hline
$N(t,v,s,b)$        & boolean to denote if there are any instances of model variant $v$ for task $t$ running on GPU segment $s$ with a maximum batch size $b$ \\
$\hat{L}(t)$        & latency of the slowest instance of task $t$ \\
$\hat{F}(t, t')$    & average multiplicative factor from task $t$ to task $t'$ \\
$\hat{R}(t)$        & demand (req/s) at task $t$ \\
$\hat{S}(t)$        & total GPU slices to run task $t$ \\
$\hat{H}(t,v,s,b)$  & throughput of an instance described by $(t,v,s,b)$ \\
$\hat{A}(t)$        & effective accuracy of task $t$ \\
$A_p$             & accuracy of path $p$ \\
$A_{obj}$         & accuracy of a candidate configuration normalized to the maximum possible accuracy \\
\hline
\multicolumn{2}{|c|}{\textbf{Decision Variables}} \\
\hline
$M(t,v,s,b)$        & number of instances of model variant $v$ for task $t$ running on GPU segment $s$ with a maximum batch size $b$ \\
\hline
\end{tabular}
}
\vspace{-0.20in}
\end{table}
\begin{equation}
\label{eq:definebinaryvarm}
  N(t,v,s,b)=\begin{cases}
    1, & \text{if $M(t,v,s,b) \geq 1$} \\
    0, & \text{otherwise}.
  \end{cases}
\end{equation}
\textit{Latency constraints.} To enforce the latency SLO, we ensure the maximum request time along any path respects the request's deadline. The worst-case latency occurs when a request encounters the slowest instance of every task in its path. $\hat{L}(t)$ calculates the per-task time using the binary variable $N(t,v,s,b)$ to find the latency of the slowest instance of $t$ in a candidate configuration.
\begin{equation}
    \label{eq:latencyonepath}
    \hat{L}(t) = \max\limits_{\mathclap{
  \substack{v \in V_t \\ s \in S \\ b \in B}
  }} ( L(t, v, s, b) * N(t, v, s, b) )
\end{equation}
Similar to prior works \cite{nexus, gpulet, loki, proteus}, we double the inference latency to account for queuing delays in forming a batch. The sum of queuing and inference times across all tasks must not exceed the latency SLO.
\begin{equation}
    \label{eq:latencyslo}
    \sum\limits_{t \in p}^{} 2*\hat{L}(t) \leq SLO_{l}, \forall p \in P
\end{equation}
\textit{Throughput constraints.} To meet throughput demands, \proj{} determines the number of instances for each task's model variants. We calculate a task's required throughput by multiplying its preceding task's request load by the task's average multiplicative factor (over the past 5 demand timestamps if the data is available), which can change across different MILP runs. $\hat{F}(t, t')$ calculates the multiplicative factor of task $t$ to $t'$ by using the active instances inferred from $N(t,v,s,b)$, to aggregate the multiplicative factor over all active model variants of task $t$.
\begin{equation}
    \label{eq:multiplicativefactoroftask}
    \hat{F}(t, t') = \sum\limits^{}_{\mathclap{
  \substack{v \in V_t \\ s \in S \\ b \in B}
  }} F(t, v, t') * N(t, v, s, b)
\end{equation}
The demand at task $t$ from all the preceding tasks $P(t)$ is $\hat{R}(t)$. For the first task in the compound inference system, $\hat{R}(t) = R$, which can change across different MILP runs. For any other task $t$, $\hat{R}(t)$ is calculated by summing the product of the multiplicative factor and the demand of all preceding tasks.
\begin{equation}
    \label{eq:workloadattaskt}
    \hat{R}(t) = \sum\limits_{t' \in P(t)}^{} \hat{R}(t') * \hat{F}(t', t)
\end{equation}
To satisfy all the requests of a task $t$, the total throughput of all the model instances must be greater than the required throughput of that task.
\begin{equation}
\label{eq:throughputrequirement}
    \sum\limits^{}_{\mathclap{
  \substack{v \in V_t \\ s \in S \\ b \in B}
  }} M(t,v,s,b) * H(t,v,s,b) \geq \hat{R}(t), \forall t \in T
\end{equation}
\textit{Total resources constraints.} To explore the tradeoff of varying the amount of computational resources allocated to each task, we add a constraint on the number of GPU slices \footnote{This might not correspond to the number of GPUs due to the placement constraints of MIG instances.} the entire compound inference system can use. 
To calculate the total number of GPU slices used, an intermediate variable $\hat{S}(t)$ denotes the total number of GPU slices used for task $t$ in the candidate configuration.
\begin{equation}
    \label{eq:resourcesfortask}
    \hat{S}(t) = \sum\limits^{}_{\mathclap{
  \substack{v \in V_t \\ s \in S \\ b \in B}
  }} M(t, v, s, b) * s_n
\end{equation}
The total resource cost in terms of GPU slices in the entire compound inference system has to be less than the total available resources.
\begin{equation}
    \label{eq:resourceslimit}
    \sum\limits_{t \in T}^{} \hat{S}(t) \leq S_{avail}
\end{equation}
\textit{Accuracy constraints.} To measure the accuracy of each path in the compound inference system, we use the pipeline accuracy score (PAS) proposed by \cite{ipa}. PAS is calculated as the product of the individual accuracy metrics of each task. We use PAS as a heuristic, and our work is orthogonal to research that finds better ways to measure the accuracy of composed ML models. The effective accuracy of a task $\hat{A}(t)$ is the weighted average of the accuracies of the candidate model variants in each instance, where the weights are the total throughput of each instance. The total throughput of an instance type $(t,v,s,b)$ is $\hat{H}(t,v,s,b)$, and the total throughput of all instance types are used to calculate the effective accuracy of a task $t$.
\begin{equation}
    \label{eq:throughputofinstance}
    \hat{H}(t,v,s,b) = M(t, v, s, b) * H(t, v, s, b)
\end{equation}
\begin{equation}
    \label{eq:accuracyfortask}
    \hat{A}(t) = \left( 
    \frac{\sum\limits^{}_{\mathclap{
  \substack{v \in V_t \\ s \in S \\ b \in B}
  }} \hat{H}(t,v,s,b) * A(t, v)}
    {\sum\limits^{}_{\mathclap{
  \substack{v \in V_t \\ s \in S \\ b \in B}
  }} \hat{H}(t,v,s,b)}
    \right)
\end{equation}
The accuracy for a path $p$ is $A_p$, and can be calculated by multiplying the accuracies of each task in the candidate configuration.
\begin{equation}
    \label{eq:pasdefinition}
    A_p = \prod\limits_{t \in p}^{} \hat{A}(t)
\end{equation}
To find the overall accuracy of the entire compound inference system, we take a convex combination \footnote{A convex combination is a weighted average where the weights are non-negative and add up to 1.} of the accuracies along each path weighted by the number of requests that take each path. We then normalize the weighted average with the maximum possible overall system accuracy ($A_{max}$) of the application. We calculate $A_{max}$ in the same way as $A_{obj}$, but we use only the most accurate model variants for Equation \ref{eq:accuracyfortask} instead of all variants $v \in V_t$.
\begin{equation}
    \label{eq:normpasdefinition}
    A_{obj} = \frac{\sum\limits^{}_{p \in P} f_p * A_p}{A_{max}}
\end{equation}
We add a constraint on the overall accuracy to be above the specified threshold, relative to the maximum overall accuracy possible for the application.
\begin{equation}
    \label{eq:accuracythreshold}
    A_{obj} \geq SLO_a
\end{equation}
{\em Objective function.} The objective of the optimization is to maximize the overall system accuracy and minimize the resources used with equations (\ref{eq:latencyslo}), (\ref{eq:throughputrequirement}), (\ref{eq:resourceslimit}), and (\ref{eq:accuracythreshold}) as constraints. The MILP maximizes the objective function described by (\ref{eq:milpobjective}) while choosing the decision variables $M(t,v,s,b)$, where $\alpha$ and $\beta$ scale the relative importance of the overall accuracy and the total resources used for a configuration.
\begin{equation}
    \label{eq:milpobjective}
    \mathop{\max_{M(t,v,s,b)}} \left( \alpha A_{obj} - \beta \sum\limits_{t \in T}^{} \hat{S}(t) \right)
\end{equation}


\subsection{Batching and Early Dropping} \label{batchingearlydropping}

The MILP finds the maximum batch size $b$ that a model instance can use. After waiting a maximum time of $\hat{L}(t)$ or if the inference of the previous batch has finished early, the model instance starts executing the inference for a new batch even if the size of the batch is less than $b$.

Similar to prior works \cite{loki}, \proj{} implements early dropping mechanisms where some requests are dropped to decrease the overall SLO violation rate. At runtime, each task can take a decision to drop a request if it is not possible to meet the latency SLO for the request even if the fastest model variants of the subsequent tasks serve the request with no delay due to batch formation. A task also drops requests that are considered to be stale, which can happen if all the model instances filled up their batches and the request is not picked up by any model instance of the task.

\section{Evaluation Methodology} \label{evaluation}

To empirically evaluate \proj{}, we implemented the system in $\sim$5K lines of Python code. We run all experiments on a Dell PowerEdge XE9640 server equipped with four NVIDIA H100 SXM 80GB GPUs. The node features 96 CPU cores, 1,007 GiB of RAM, and a configurable storage capacity. Our software environment includes Ubuntu 22.04, Python 3.10, CUDA 12.4, and PyTorch 2.6.0. We use the \lstinline{nvidia-mig-parted}\footnote{https://github.com/NVIDIA/mig-parted} tool to set up the MIG instances on all GPUs, and Gurobi \cite{gurobi} to find solutions to the MILP. We provide more details on the parameters chosen for the evaluation in \S \ref{parameterschosen}, and the implementation of the compound inference system in Appendix \ref{implementation}.

\subsection{Workloads}

\begin{figure*}[t]
    \centering
    \begin{subfigure}[t]{0.32\textwidth}
        \centering
        \includegraphics[width=1.0\textwidth,page=1]{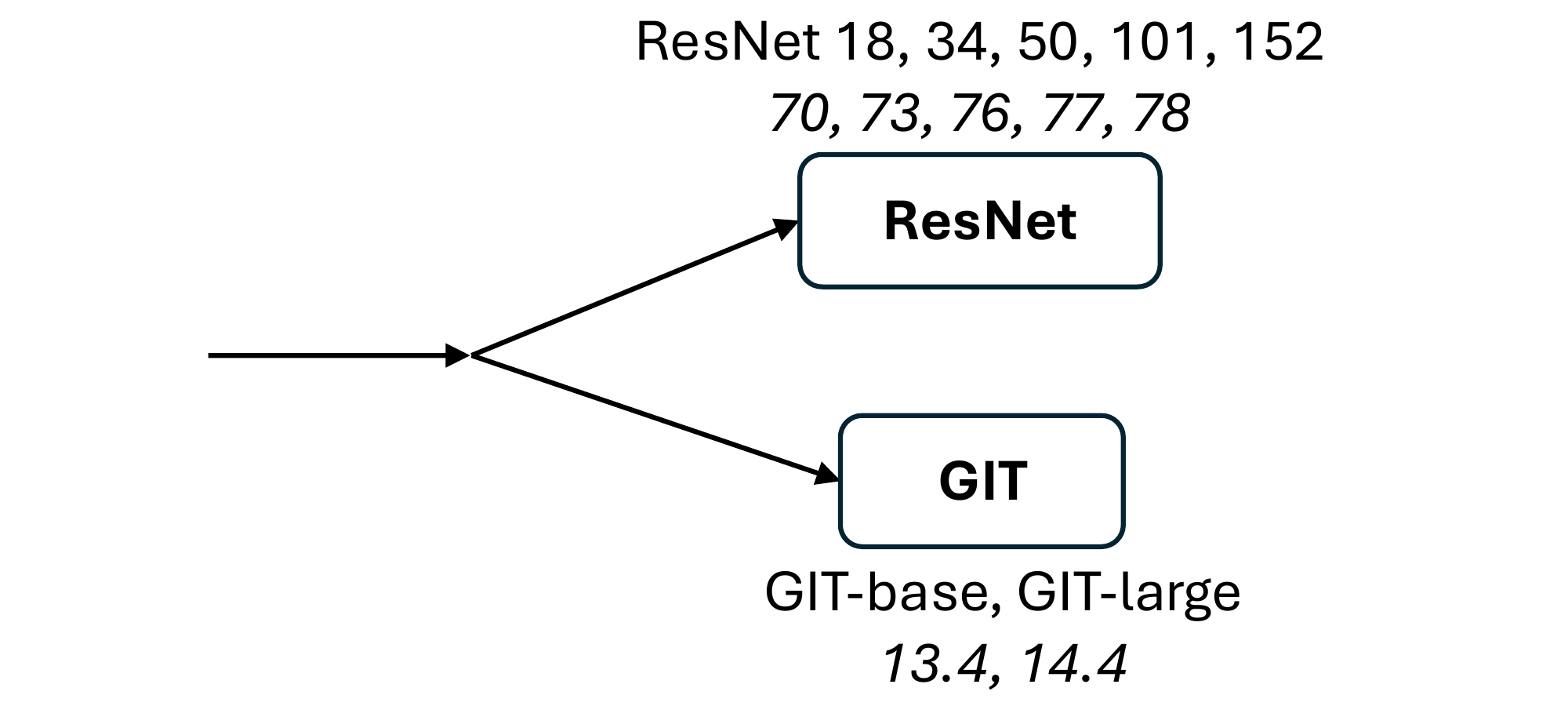}
        \caption{Social media application.}
    \end{subfigure}
    ~
    \begin{subfigure}[t]{0.32\textwidth}
        \centering
        \includegraphics[width=1.0\textwidth,page=2]{figures/evaluationapps.pdf}
        \caption{Traffic analysis application.}
    \end{subfigure}%
    ~ 
    \begin{subfigure}[t]{0.32\textwidth}
        \centering
        \includegraphics[width=1.0\textwidth,page=3]{figures/evaluationapps.pdf}
        \caption{AR assistant application.}
    \end{subfigure}
    \caption{Applications and their inference task graphs used to evaluate \proj{}. Each task is annotated with the model variants that can perform the task, along with the relevant accuracy metrics.}
    \label{fig:appstaskgraphs}
    \vspace{-0.20in}
\end{figure*}

We evaluate \proj{} with three applications, built as compound inference systems with different depths of the paths a request can take. These applications consist of a diverse range of ML models in terms of their functionality, number of parameters, and resource requirements. 
The tasks of the applications and the model variants \proj{} considered for each task are summarized in Figure \ref{fig:appstaskgraphs}.
\noindent
\begin{enumerate}[topsep=1pt,noitemsep, leftmargin=1em]
    \item \textit{Social media} \cite{loki} (depth of 1): Each input image is processed by two models for the social media application. A ResNet model \cite{resnet} predicts the object present in the image, and a GIT model \cite{microsoftgit} generates a caption for the image.
    \item \textit{Traffic analysis} \cite{loki} (depth of 2): Each input image is processed by a YOLO \cite{yolo} model to detect if there are any cars or people in the image. For each detected car, an EfficientNet model \cite{efficientnet} predicts the make and model of the car. For each detected person, a VGG model \cite{vgg} predicts the gender and age of the person.
    \item \textit{AR assistant} (depth of 3): Each input image is first processed by a YOLO model \cite{yolo} to detect an object. A GIT model \cite{microsoftgit} generates a caption to describe the object, and a text-to-speech (TTS) model \cite{vits,glowtts} converts the textual description of the object into an audio description.
\end{enumerate}

When performing accuracy scaling, the accuracy values we use for each model variant are the relevant metrics reported in the available public repositories \cite{yolov5metrics, resnetmetrics, vggmetrics} and the manuscripts \cite{vits, microsoftgit, efficientnet} of the model architecture families.

To evaluate the configurations generated by \proj{} at different demand (req/s) values that a serving system can observe in an entire day, we obtain values for the demand over a day using the request timing information from a Twitter trace \cite{twittertrace, twitterfrankdonnelly}. We bin the requests into 5-minute intervals, where each \textit{demand timestamp} in the binned demand trace represents the average demand over each 5-minute interval throughout the day from the Twitter trace. The trace contains a diurnal pattern that is representative of production serving demand \cite{markcloudinferenceserving, azureserverlessworkload}. We scale the entire trace to the maximum demand that can be served by \proj{} for each application, while preserving the broad trends within the trace. 

For the payload of the requests, we use the COCO dataset \cite{cocodataset} for the social media and AR assistant applications, and the Bellevue traffic dataset \cite{bellevuetrafficvideodataset} for the traffic analysis application.

\subsection{MILP and Reconfiguration Frequency}

To test the efficacy of \proj{} at different demand conditions a serving system can observe in a day, we invoke the MILP and consequent system reconfiguration at each demand timestamp bin of the trace described above. We use a rudimentary demand prediction approach to average the demand of the 5 previous timestamp bins with an additional slack. We run our complete system for a short duration for each timestamp bin until it reaches a steady state. While the MILP and system reconfiguration incur overheads, these are negligible relative to the frequency at which they are invoked. The MILP in particular can be invoked in the background and takes 2s to 20s, depending on the demand conditions and application.

\subsection{Baselines Evaluated}

We point out the configuration search spaces that the baselines consider when they perform accuracy scaling (A), enable spatial partitioning (S), and perform task-graph-informed resource budgeting (T).

For the analytical evaluation, we compare the maximum serviceable demand of \proj{} with baselines that contain all combinations of A/S/T in their configuration search space. For the empirical evaluation, we compare \proj{} against baselines whose configuration search spaces perform task-graph-informed resource budgeting along with A or S. We also compare against A+S, which is task-graph-uninformed and does not exist in prior work but performs well in our analytical evaluation. Some of these baselines are equivalent or close to the configuration search spaces of relevant prior works Loki \cite{loki}, Clover \cite{clover}, and ParvaGPU \cite{parvagpu}.


Loki performs accuracy scaling for each task in a task-graph-informed manner, and is equivalent to A+T. ParvaGPU+T is equivalent to S+T since it enables spatial partitioning and is task-graph-informed and does not perform accuracy scaling. Clover performs accuracy scaling and enables spatial partitioning in a task-graph-uninformed manner, but does not use MPS within the MIG instances to increase the utilization. Clover+MPS correctly describes A+S, since A+S will perform at least as well as Clover due to better utilization of the GPUs. Loki is the only previous work that represents a configuration space of using any two of the three features from Table \ref{tab:featuresrelatedworks}. \proj{}'s configuration search space is A+S+T, and is the superset of all relevant prior works.

Baselines that do not perform accuracy scaling consider only the highest‑accuracy variant, and those that do not enable spatial partitioning exclusively use whole‑GPUs for each model. For task‑graph‑uninformed baselines we detail a static task-wise budget allocation method in Appendix \ref{taskgraphuninformed} that makes them as strong as possible.

\subsection{Parameters for Evaluation} \label{parameterschosen}

When finding a configuration for a demand timestamp, we use a slack of 5\% similar to previous inference serving systems \cite{infaas}. We use $\alpha=1$ and $\beta=0.035$ for Equation \ref{eq:milpobjective} to scale the range of available GPU slices ($0-28$) in our testbed to the range of the overall system accuracy ($0-1$) and give equal importance to both objectives. For the accuracy SLO, we allow \proj{} to decrease the overall system accuracy until a threshold of 90\% of the maximum possible system accuracy for each application. We use latency SLOs of 650ms, 700ms, and 1550ms for the traffic analysis, social media, and AR assistant applications to make sure that all configuration spaces would be able to serve the applications at some demand. When running the end-to-end system, we account for the communication latency by adding the time for each communication hop ($\sim$10ms) to the latency SLO based on the depth of the application. To measure the staleness of requests, we use values of 20ms for the social media and traffic analysis applications, and 40ms for the AR assistant application in proportion to their end-to-end latency SLOs.

\subsection{Metrics}

We present three metrics for each baseline in the empirical evaluation: cost in terms of percentage of available GPU slices used, overall system accuracy drop, and the latency SLO violation rate. We define the overall system accuracy drop as a percentage of the maximum overall accuracy possible for an application. We define the latency SLO violation rate (also referred to as the SLO violation rate for brevity) as the ratio of number of requests that miss their deadline to the total number of requests, similar to previous works that consider compound inference systems \cite{loki}. We highlight that a baseline performs particularly poorly if the SLO violation rate is $\geq$ 10\%. We also consider an early dropped request as a deadline violation, and multiple violations if the task performing the early dropping has a multiplicative factor greater than 1.

\section{Results} \label{results}

We first analyze the solutions of the MILP to find the maximum throughputs that can be served with each configuration search space. We then run empirical evaluations of the end-to-end compound inference systems with the generated configurations to report the metrics, and analyze the configurations chosen by \proj{}. 

\begin{figure}[htbp]
  \includegraphics[width=0.48\textwidth]{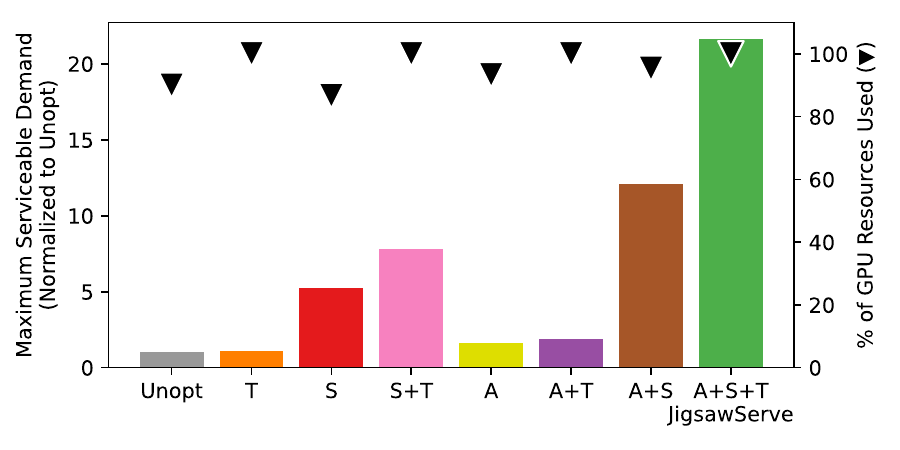}
  \caption[Maximum demand for the various configuration search spaces.]{Maximum demand that can be satisfied for the traffic analysis application on a large testbed by exploring combinations of A/S/T. Higher maximum demand is better.
  }
  \label{fig:maxdemands}
    
\end{figure}

\textbf{Maximum serviceable demand.} Figure \ref{fig:maxdemands} shows the maximum possible demand that can be served for the traffic analysis application to compare scenarios where a serving system performs accuracy scaling (A) while keeping the overall system accuracy above a threshold of 90\%, enables spatial partitioning (S), and does task-graph-informed resource budgeting (T). We choose to run the MILP for a hypothetical testbed of 120 GPUs (840 GPU slices) to ensure that all configurations would be able to exactly follow any task-wise resource budgets. Figure \ref{fig:maxdemands} shows the maximum serviceable demand of each configuration search space normalized to that of an unoptimized system, along with the actual percentage of GPU slices that each configuration search space uses to achieve its highest serviceable demand.

When the serving system is unoptimized (Unopt), it does not perform accuracy scaling, disables spatial partitioning, and does task-graph-uninformed resource budgeting. Unopt has the lowest maximum serviceable demand among all other configuration search spaces. Unopt is also unable to use all the available GPUs due to the task-wise resource budgets since it is task-graph-uninformed. 
T does task-graph-informed budgeting, resulting in a maximum serviceable demand that is $1.1\times$ that of Unopt while being able to use all the available GPUs. All other task-graph-uninformed configuration search spaces (S, A, A+S) similarly do not use all available GPUs, or GPU slices if they enable spatial partitioning.

S enables spatial partitioning of GPUs enabling fine-grained resource distribution among tasks, and achieves a maximum serviceable demand that is $5.25\times$ the Unopt version. S delivers the highest standalone gain over Unopt in serving capacity for the same GPU resources, compared to A (1.6$\times$) and T (1.1$\times$).

A+T and S+T achieve a higher maximum serviceable demand that is $1.9\times$ and $7.8\times$ respectively while using all the available GPUs. A+S achieves a maximum serviceable demand that is $12\times$, which is a great increase compared to all the other configuration search spaces that use either one or two of A, S, and T. A+S+T corresponds to \proj{}, and achieves the highest maximum serviceable demand that is $21.6\times$ that of Unopt. A+T is closest to (and explores a larger search space than) prior work of Loki \cite{loki}, and \proj{}'s maximum serviceable demand is 11.3$\times$ that of A+T. This demonstrates how performing accuracy scaling by choosing the model variants for each task in the compound inference system, enabling spatial partitioning on GPUs, and performing task-graph-informed resource budgeting allows \proj{} to serve a much higher demand using the same GPU resources. 



\begin{figure*}[t!]
    \centering
    \begin{subfigure}[t]{0.32\textwidth}
        \centering
        \includegraphics[width=\textwidth]{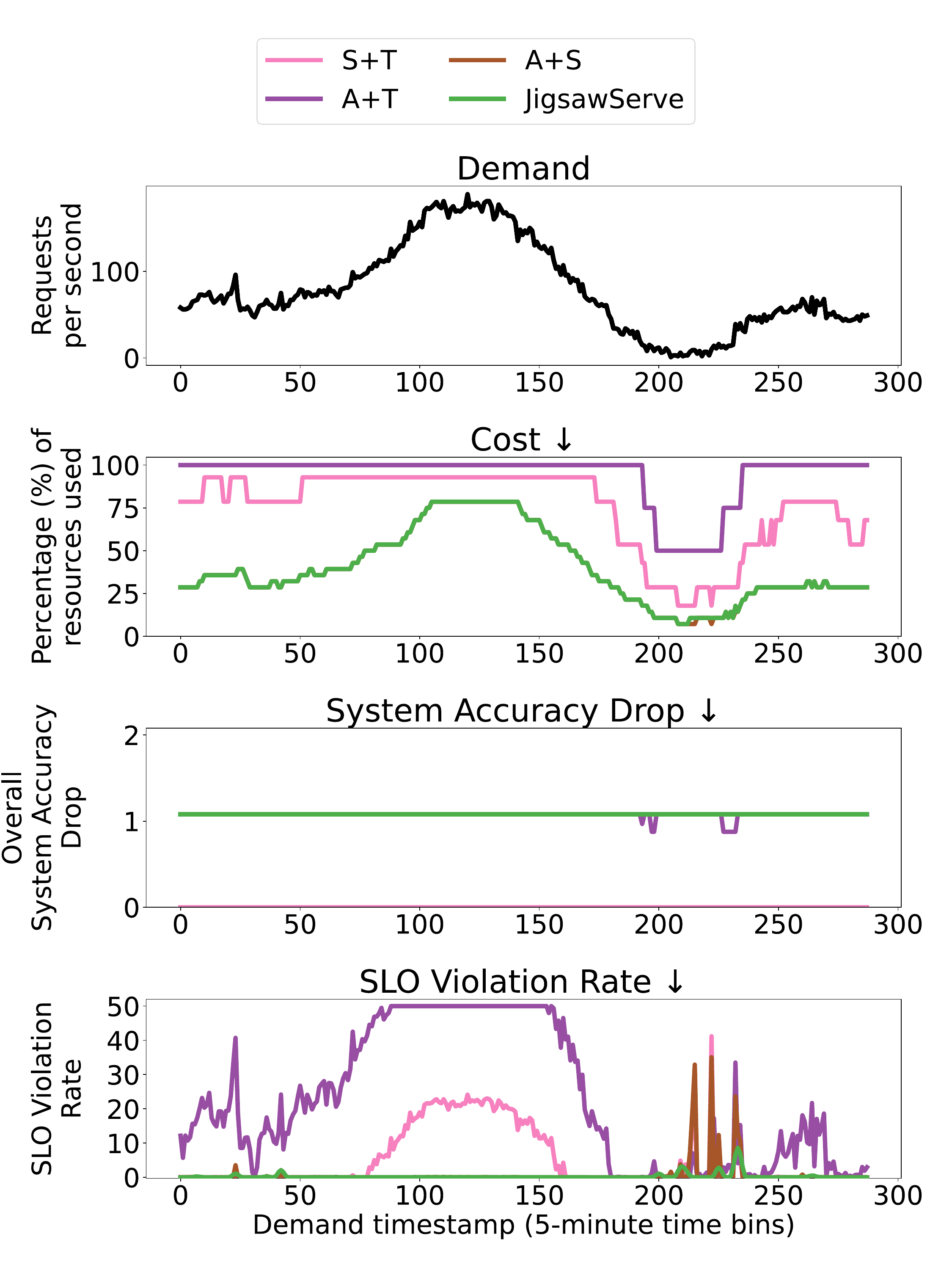}
        \caption[Results for the social media application.]{Social media application.}
        \label{fig:socialmediaappresults}
    \end{subfigure}%
    \hfill
    \begin{subfigure}[t]{0.32\textwidth}
        \centering
        \includegraphics[width=\textwidth]{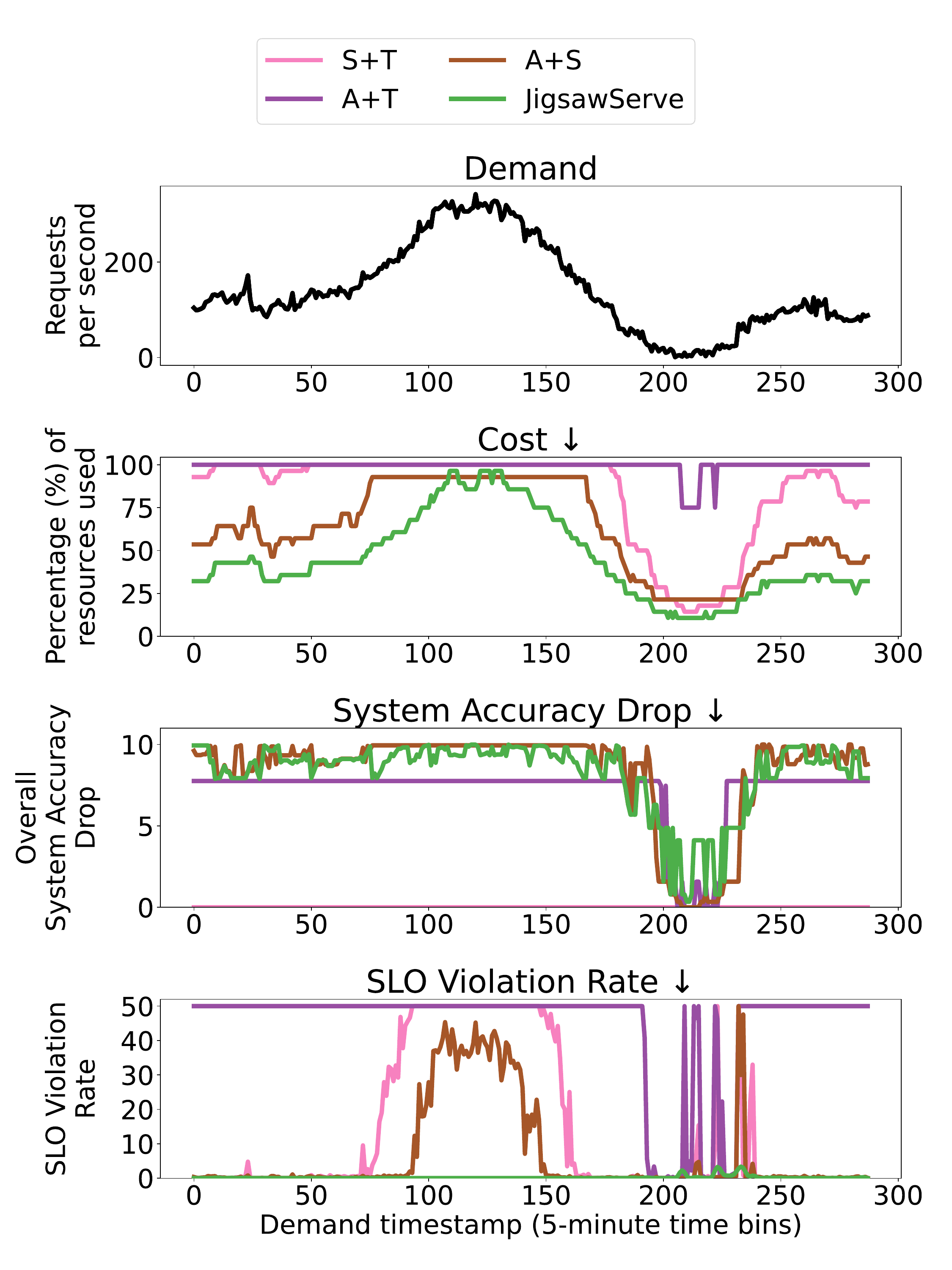}
        \caption[Results for the traffic analysis application.]{Traffic analysis application.}
        \label{fig:trafficanalysisappresults}
    \end{subfigure}
    \hfill
    \begin{subfigure}[t]{0.32\textwidth}
        \centering
        \includegraphics[width=\textwidth]{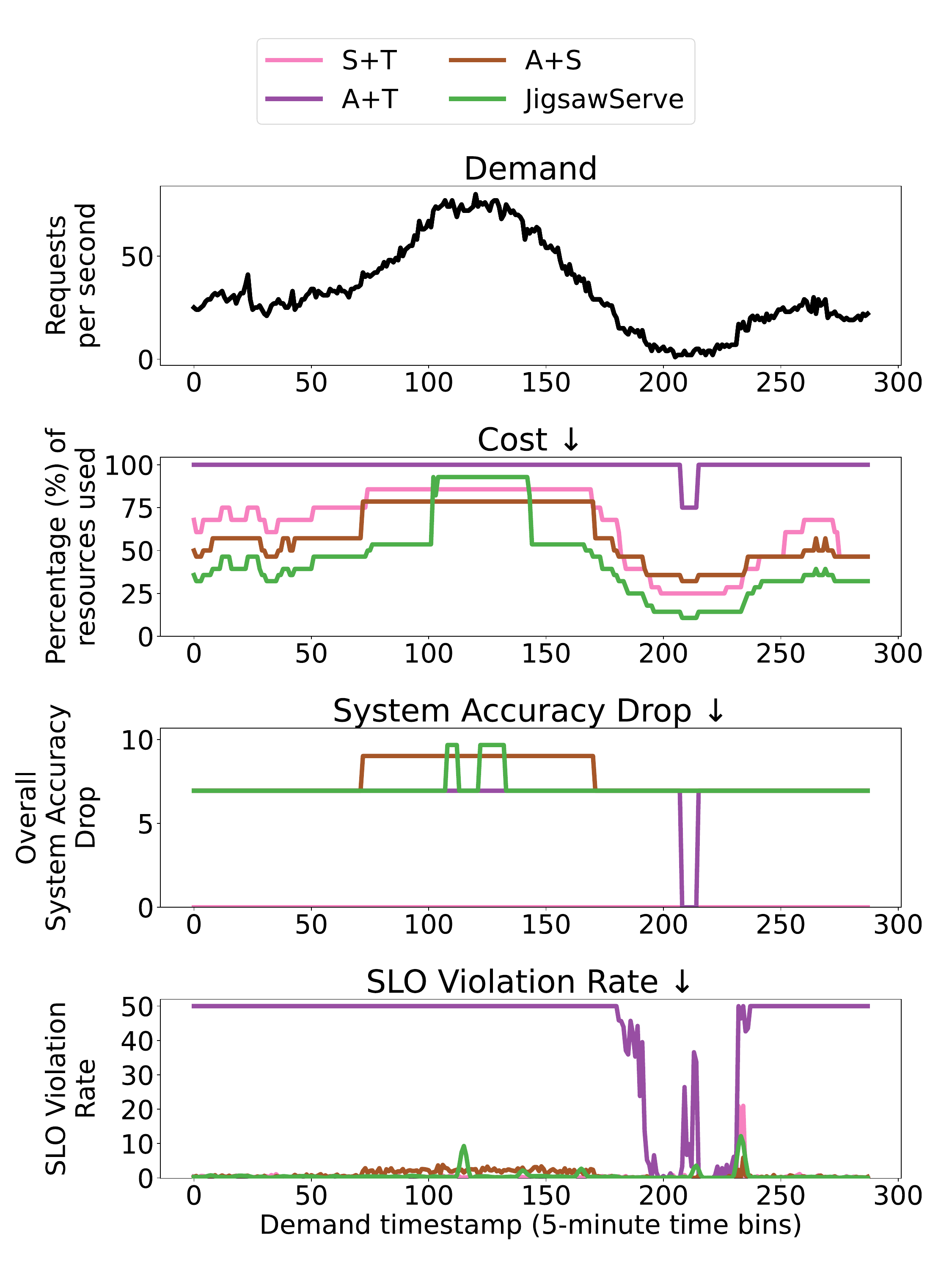}
        \caption[Results for the AR assistant application.]{AR assistant application.}
        \label{fig:arassistantappresults}
    \end{subfigure}
    \hfill   
    


    \begin{subfigure}[t]{0.32\textwidth}
        \centering
        \includegraphics[width=\textwidth]{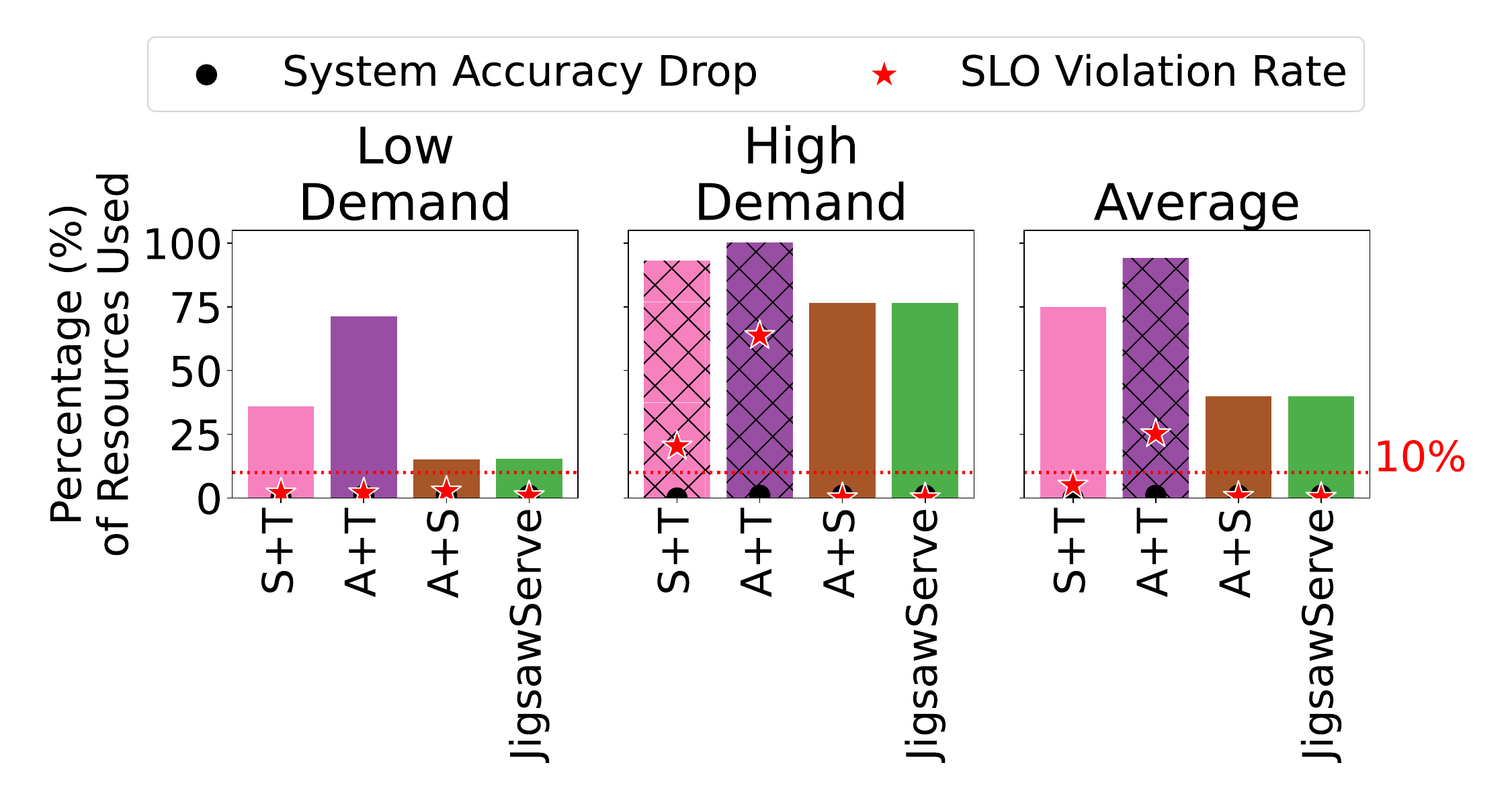}
        \caption[Results for the social media application.]{Social media application.}
        \label{fig:socialmediaappatdemand}
    \end{subfigure}%
    \hfill
    \begin{subfigure}[t]{0.32\textwidth}
        \centering
        \includegraphics[width=\textwidth]{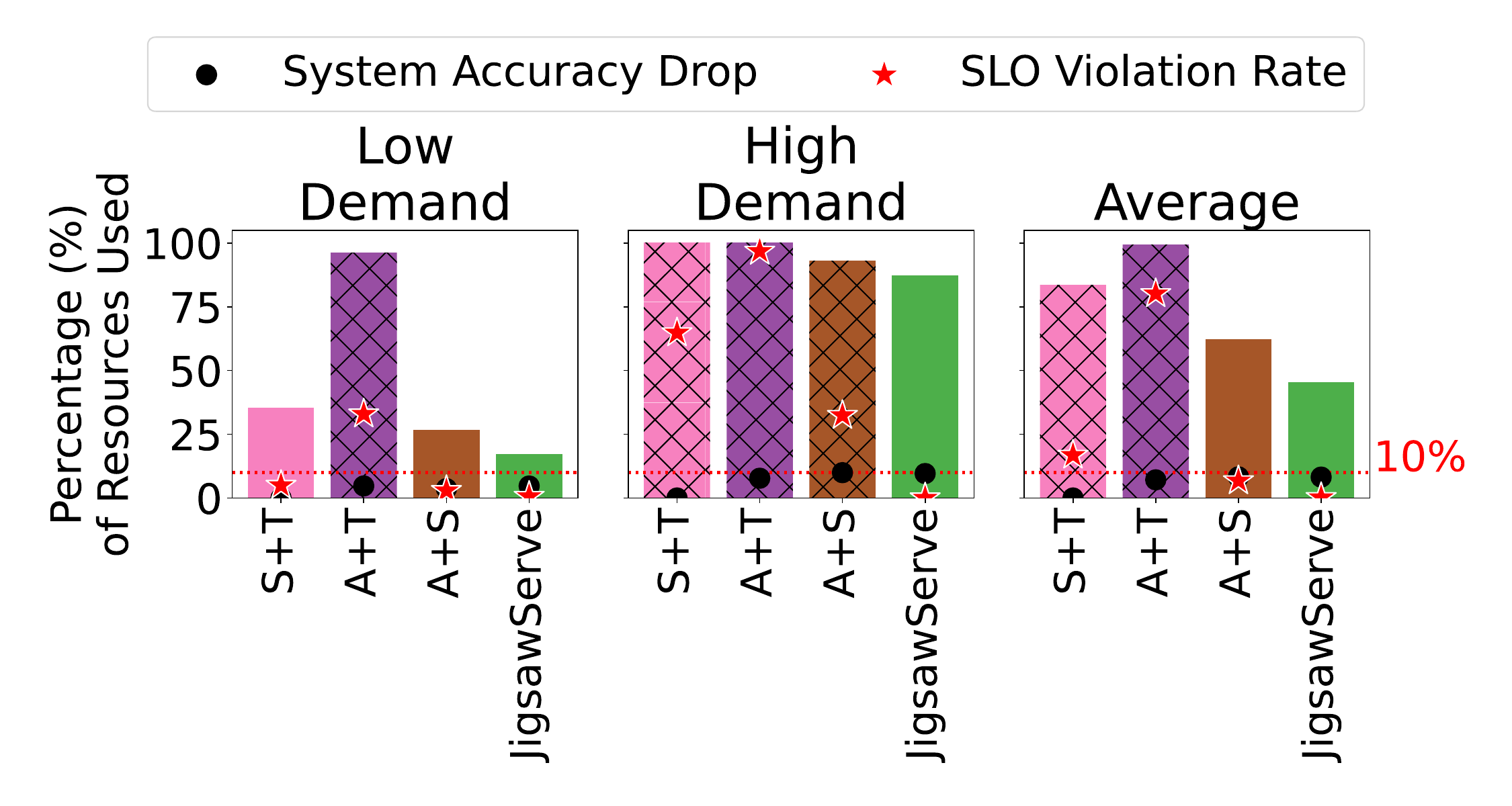}
        \caption[Results for the traffic analysis application.]{Traffic analysis application.}
        \label{fig:trafficanalysisappatdemand}
    \end{subfigure}
    \hfill
    \begin{subfigure}[t]{0.32\textwidth}
        \centering
        \includegraphics[width=\textwidth]{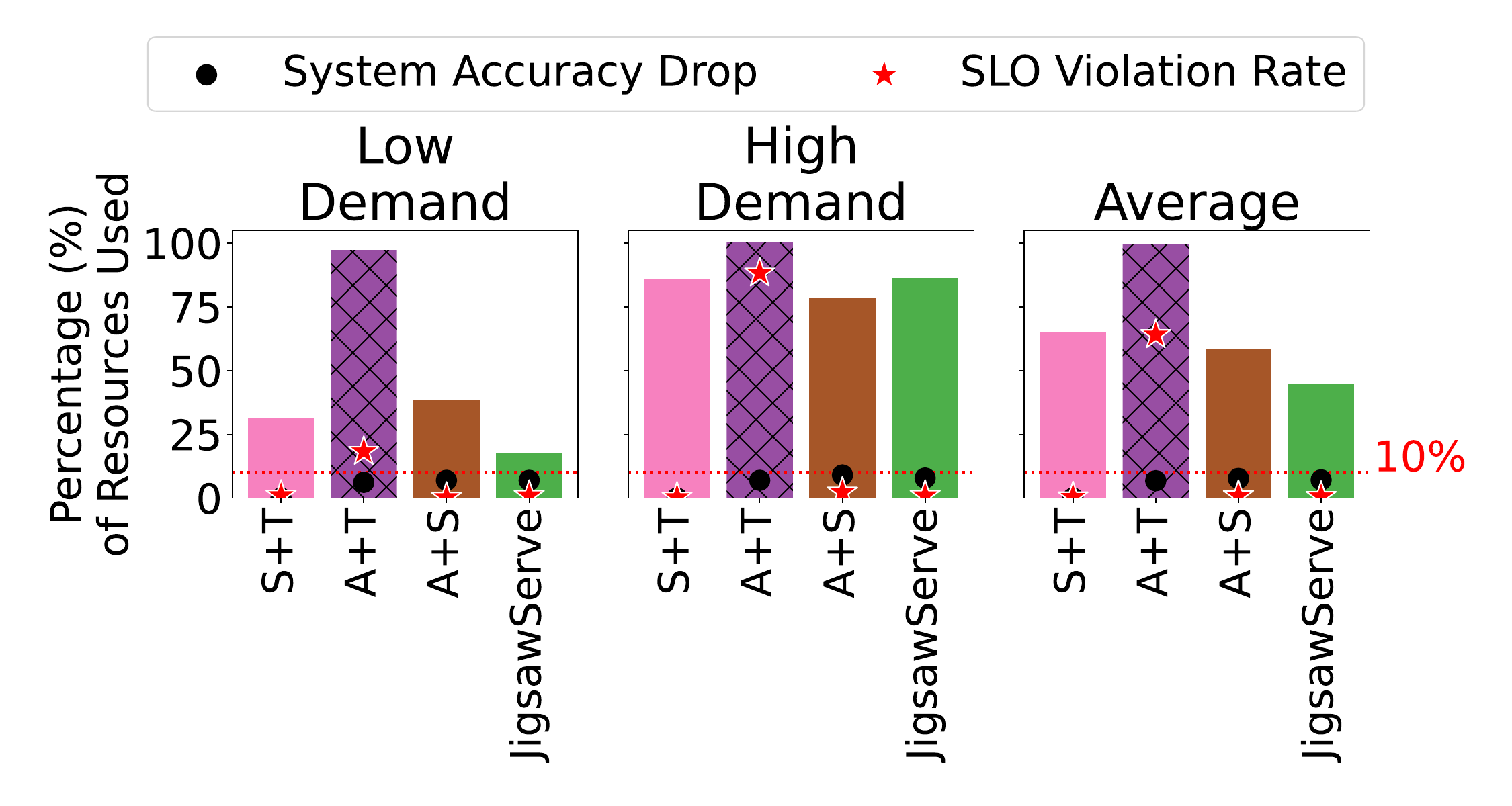}
        \caption[Results for the AR assistant application.]{AR assistant application.}
        \label{fig:arassistantappatdemand}
    \end{subfigure}
    \hfill   
    \caption{(a), (b), and (c) show the statistics of all demand timestamps for the evaluated applications. We cap the SLO violation rate at 50\% for better visualization. (d), (e), (f) show the aggregated statistics for the low demand conditions (timestamp 180-240), high demand conditions (timestamp 100-150), and the average over all demand timestamps. The bars are hatched if the SLO violation rate is $\geq$10\%. For all metrics, lower is better.}
    \label{fig:variousdemandresults}
    \label{fig:allresults}
\end{figure*}

\textbf{Empirical evaluation of the end-to-end system.} 
We empirically evaluate the top four performing systems identified through our comprehensive analytical evaluation that use at least two features from Table \ref{tab:featuresrelatedworks}: S+T, A+T, A+S, and \proj{}. Figure \ref{fig:allresults} shows the demand, resource consumption, accuracy drop, and SLO violations when \proj{} and the selected baselines find configurations to serve various demands for the evaluated applications. 

\proj{} and A+S, both use only 40\% of the available GPU slices on average across all demand timestamps for the social media application. The configurations chosen by them are also identical, since the social media application only has a depth of 1 and does not benefit from task-graph-informed resource budgeting. For the traffic analysis (depth 2) and AR assistant (depth 3) applications, \proj{} needs less GPU slices when compared to A+S and highlights the advantage of performing task-graph-informed budgeting of resources. A+S also results in an SLO violation rate that is $\geq$10\% at high demand conditions for the traffic analysis application even though it uses less GPU slices than A+T and S+T.

When comparing \proj{} and S+T, \proj{} uses fewer GPU slices across all applications. For the social media and traffic analysis applications, S+T is unable to serve a high demand and results in $\geq$10\% SLO violation rate. 
A+T performs the worst among all the baselines, and has $\geq$10\% SLO violation rate even at low demand conditions in two of three applications. Since S+T, A+T, and \proj{} are task-graph-informed, they are able to make use of all available resources at high demand conditions if necessary, unlike A+S. 

\textbf{Broad trends across all applications and baselines.} All the baselines are able to find valid configurations at low demand conditions, which can be seen by the relative low SLO violation rate. When any baseline is unable to find a valid configuration to solve a demand, it uses the configuration that can serve the highest demand. This results in the percentage of resources used plateauing at its highest value during high demand conditions for all baselines. At low demand conditions, even though all baselines are able to find valid configurations, they are sensitive to fluctuations in the demand and result in spikes in the SLO violation rate. This happens when the actual demand of a timestamp is greater than the demand predicted from the previous timestamps. A difference of 5 req/s appears small in the demand graph, but can be a 2$\times$ increase if a baseline finds a configuration for 5 req/s but actually observes a demand of 10 req/s.

\proj{} shows the best overall performance, using just 43.3\% of the available GPU resources on average, respecting the accuracy SLO, and showing less than 0.6\% average SLO violations. In contrast, S+T and A+T show over 10\% SLO violations and use more than 2$\times$ the resources of \proj{} in at least one case. A+S reaches 6.7\% SLO violations and 33\% more resources than \proj{} in at least one case.

\begin{figure}[t!]
    \centering
    \begin{subfigure}[t]{0.32\textwidth}
        \centering
        \includegraphics[width=\textwidth]{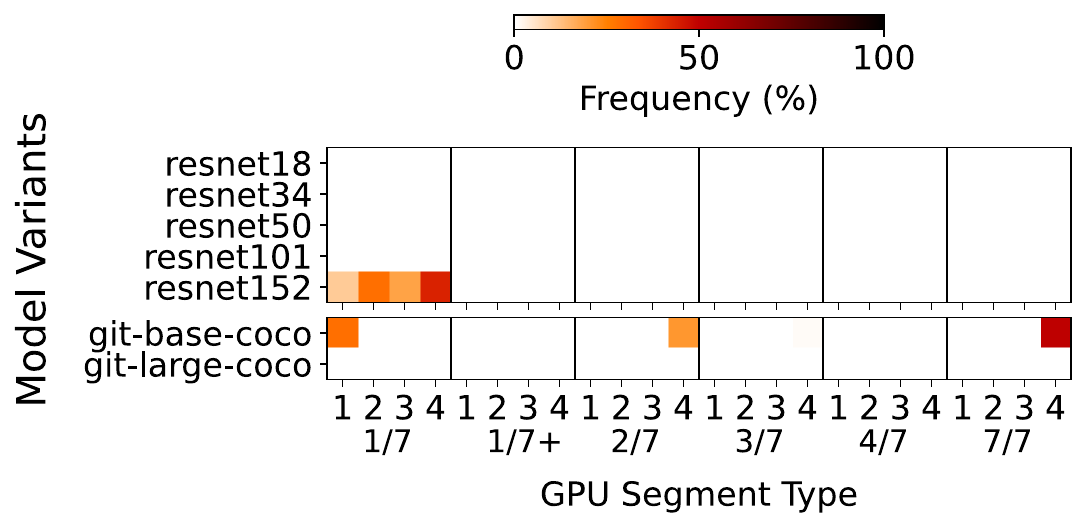}
        \caption[Configurations chosen for the social media application.]{Social media application.}
        \label{fig:socialmediaappconfigs}
    \end{subfigure}%
    \hfill
    \begin{subfigure}[t]{0.32\textwidth}
        \centering
        \includegraphics[width=\textwidth]{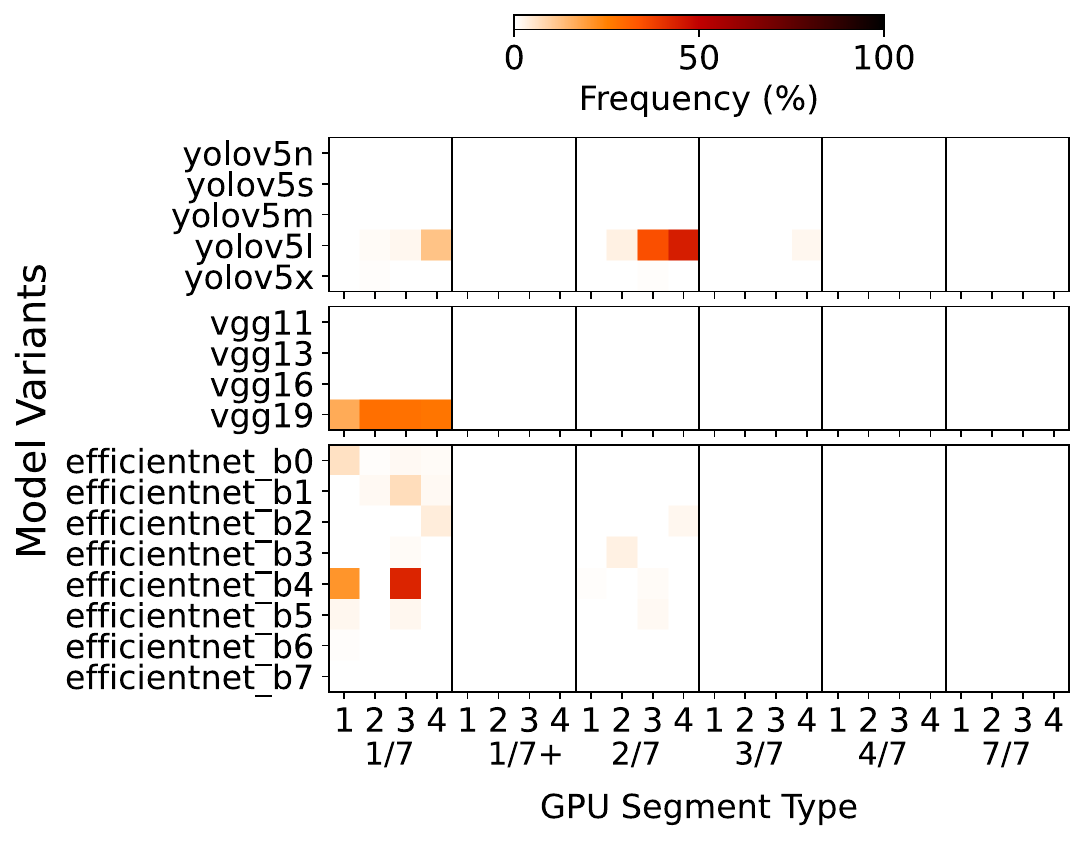}
        \caption[Configurations chosen for the traffic analysis application.]{Traffic analysis application.}
        \label{fig:trafficanalysisappconfigs}
    \end{subfigure}
    \hfill
    \begin{subfigure}[t]{0.32\textwidth}
        \centering
        \includegraphics[width=\textwidth]{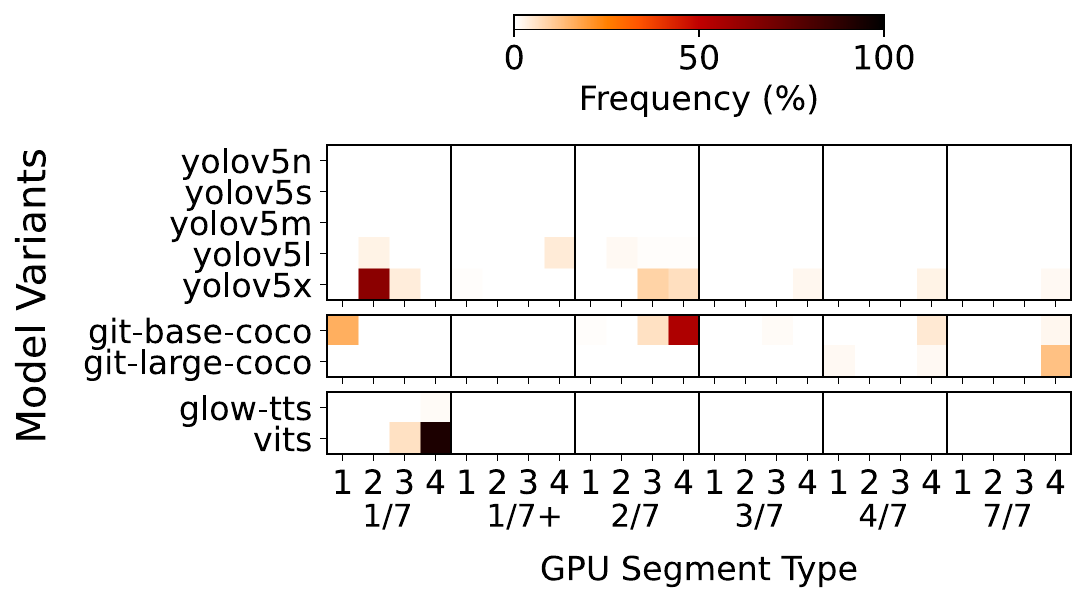}
        \caption[Configurations chosen for the AR assistant application.]{AR assistant application.}
        \label{fig:arassistantappconfigs}
    \end{subfigure}
    \hfill   
    \caption[Analyzing chosen configurations.]{The frequency of model variants and GPU segment types in the configurations chosen by \proj{} for the evaluated applications. The GPU segment types are described by both a MIG instance type (1/7 - 7/7) and an MPS concurrency level (1, 2, 3, 4).}
    \label{fig:exploringconfigs}
\end{figure}

\textbf{Analyzing the configurations chosen by \proj{}}. Figure \ref{fig:exploringconfigs} shows the model variants and GPU segments \proj{} chooses for each task in the evaluated applications. When performing accuracy scaling, \proj{} is able to choose to scale down the accuracy for tasks that will have the greatest impact on the resources required based on the application. For example, \proj{} chooses to run the yolov5l variant of YOLO for the traffic analysis, but chooses the yolov5x variant of YOLO for the AR assistant application. For the AR assistant application, \proj{} chooses to run less accurate model variant for the GIT task, while using the high accuracy models for the YOLO and TTS tasks. For the traffic analysis application, \proj{} typically chooses a mix of model variants for the EfficientNet task to keep the overall accuracy above the accuracy threshold. 

\proj{} also identifies the most efficient GPU segment type to run each model. VGG19 always uses the GPU segments described by a 1/7 MIG instance with varying MPS concurrency. In both the social media and the AR assistant application, the GIT-base model frequently uses the GPU segments described by a 1/7 MIG instance with an MPS concurrency of 1, and the GPU segments described by a 2/7 MIG instance with an MPS concurrency of 4. \proj{} chooses to run all ResNet model variants exclusively on a 1/7 MIG instance type with varying MPS concurrency. The EfficientNet model variants are frequently run on a 1/7 MIG instance with varying MPS concurrency, while occasionally using the 2/7 MIG instance type. \proj{} almost never uses the 1/7+extra memory and 3/7 MIG instance types, since they are just as expensive to use as a 2/7 or 4/7 MIG instance respectively due to their placement constraints on a GPU.

\subsection{Overhead of \proj{}}

The profiler takes 7-12 hours to exhaustively profile the entire configuration search space of all model variants, MIG instance sizes, MPS concurrency, and batch sizes. This profiling happens before the system starts serving, and has to be done only once. The time spent doing the profiling is compensated by the higher throughput and lower cost achieved when \proj{} uses the profiled information to serve the compound inference system.

The average time taken by the MILP in the controller to find the optimal configuration varies from 2-20 seconds on our testbed based on the demand and application. Finding the optimal configuration is not on the critical path of serving requests, since the controller can run the MILP asynchronously on the CPU. 
Repartitioning the MIG instances on GPUs would be another overhead due to its disruptive nature, since the serving system cannot use the GPUs while they are being repartitioned. Previous works \cite{ibmmigpacking} provide solutions for non-disruptive repartitioning by making use of additional GPUs during the repartitioning period, which is feasible when \proj{} is deployed on large clusters of datacenter GPUs.

\section{Related Works} \label{relatedworks}

There exists extensive research on serving systems for inference workloads. Most relevant to our work is research on compound inference systems \cite{nexus, ipa, slopt, loki, dream}, adaptation techniques to perform accuracy scaling \cite{loki, proteus, infaas, dream, compoundaimodelselection, clover}, and fine-grained hardware resource allocation with spatially partitioned GPUs \cite{miger, gpulet, misompsformig, parvagpu, fluidfaas, esg, migserving}. Table~\ref{tab:featuresrelatedworks} summarizes the most closely related works. 
Prior works either do not consider compound systems or do not explore spatial partitioning schemes to address datacenter GPU under-utilization, among other limitations when compared to \proj{} (\S\ref{introduction}). As we point out in \S \ref{evaluation}, \proj{}'s configuration search space is a superset of prior works, and outperforms all of them.

\textbf{Model variants for inference serving.} INFaaS \cite{infaas} and Proteus \cite{proteus} perform accuracy scaling to choose model variants to serve inference requests for individual models. Loki \cite{loki} and IPA \cite{ipa} are serving systems for compound inference workloads that select model variants for each task in the system. Loki was designed and evaluated for consumer-grade GPUs, and will result in GPU under-utilization when it runs on datacenter GPUs since it allocates an entire GPU to each model instance. \proj{} focuses on serving compound inference systems on datacenter GPUs, and co-locates models and share the GPU to improve utilization of the GPUs unlike Loki. IPA runs the individual models of the compound inference system on CPU cores and does not use GPUs to accelerate the inference. 
DREAM \cite{dream} is a scheduler that dynamically chooses the best accelerator to run each model in an XR-based compound inference for a system with multiple heterogeneous accelerators for a single user. LLMSelector \cite{compoundaimodelselection} selects the large-language models (LLMs) to use in a compound inference system with multiple LLMs interacting as agents with an objective to maximize the end-to-end system accuracy, but does not address the hardware cost, latency, or throughput of serving their workload unlike \proj{}.



\textbf{Serving systems with GPU sharing.} GPUlet \cite{gpulet} and GSLICE \cite{gslice} serve compound inference systems and share GPUs by using MPS. INFaaS shares GPUs by co-locating models with an interference-prone mechanism, and constantly probes the models to identify when the interference is too high. Clover \cite{clover} is a framework that hosts multiple model variants on different MIG instances to minimize the carbon emissions of serving inference for individual ML models, but does not consider compound inference systems. MIG-serving \cite{migserving} finds deployments of MIG instances to serve inference workloads of multiple individual models when given the throughput and latency requirements. ParvaGPU \cite{parvagpu} uses MPS-enabled MIG instance to serve individual models while maintaining a high GPU utilization. 


\section{Future Work and Conclusion} \label{discussionconclusion}

We plan to explore \proj{}'s efficacy when serving real applications, where relaxing the accuracy threshold results in a measurable and quantifiable decrease in the quality of experience. It would also be interesting to explore how the core ideas of \proj{} can be applied to more compound inference systems, beyond those whose task-graphs can be described as a DAG. Exploring more compound inference systems could involve considering large ML models that need more than 1 GPU for a single instance. Designing a serving system for such systems would require better understanding of the communication overheads and memory bottlenecks to satisfy the data dependencies.

\proj{} uses spatial partitioning mechanisms on NVIDIA GPUs, but the core ideas of \proj{} are applicable even to GPUs from other vendors that provide spatial partitioning mechanisms such as the Modular Chiplet Platform on AMD GPUs. Techniques like dynamic rerouting \cite{loki} of individual requests to model instances with unutilized capacities are complementary to our work, and will further improve the runtime performance of \proj{}.
\proj{} uses a rudimentary workload prediction to predict the future demand. Using better workload prediction methods to predict the demand well ahead of time can help hide the GPU repartitioning time during real deployment on a cluster of datacenter GPUs.

\proj{}'s MILP formulation allows for any optimization objective when exploring the configuration space of choosing model variants and MIG instance sizes for a compound inference system. For example, it can be used to minimize the carbon intensity of serving compound inference systems, similar to \cite{clover}. \proj{} currently uses a bin-packing algorithm for MIG instance placement, which can lead to GPU fragmentation. Future work can optimize for ideal MIG packing and placement as a part of the objective. Alternatively, a multi-level resource allocator can allocate pools of resources, and model instances can flexibly make real-time decisions at runtime to use from the resource pools.


\proj{} is the first serving system that jointly optimizes model variant selection and GPU spatial partitioning for compound inference systems on datacenter GPUs. By allocating resources and accuracy budgets in a task-graph–informed manner, it meets end-to-end latency and accuracy SLOs while using only 43.3\% of available GPU capacity on average. Our results demonstrate that joint optimization across accuracy scaling, spatial partitioning, and task graph structure is key to efficient serving of compound inference systems. \proj{} makes a case for two critical advances: (1) The ML community should continue to release multiple model variants per architecture to allow accuracy-performance flexibility, and (2) GPU vendors must embrace and prioritize developing spatial partitioning mechanisms.

\section*{Acknowledgments} \label{acknowledgements}
This work is supported in part by U.S. National Science Foundation grant \#2217144, the AMD Center of Excellence at UIUC, and the IBM-Illinois Discovery Accelerator Institute. This work used JetStream2 \cite{jetstream2} at Indiana University through allocation CIS250170 from the Advanced Cyberinfrastructure Coordination Ecosystem: Services \& Support (ACCESS) program \cite{access}, which is supported by U.S. National Science Foundation grants \#2138259, \#2138286, \#2138307, \#2137603, and \#2138296.

\bibliography{references}

\begin{thebibliography}{45}
\providecommand{\natexlab}[1]{#1}
\providecommand{\url}[1]{\texttt{#1}}
\expandafter\ifx\csname urlstyle\endcsname\relax
  \providecommand{\doi}[1]{doi: #1}\else
  \providecommand{\doi}{doi: \begingroup \urlstyle{rm}\Url}\fi

\bibitem[res(2017)]{resnetmetrics}
Pytorch model hub, resnet, 2017.
\newblock URL \url{https://pytorch.org/hub/pytorch_vision_resnet/}.

\bibitem[vgg(2017)]{vggmetrics}
Pytorch model hub, vgg-nets, 2017.
\newblock URL \url{https://pytorch.org/hub/pytorch_vision_vgg/}.

\bibitem[twi(2022)]{twittertrace}
Twitter streaming traces, 2022.
\newblock URL \url{https://archive.org/details/archiveteam-twitter-stream-2022-11}.

\bibitem[yol(2024)]{yolov5metrics}
Yolov5 vs. yolov8: A detailed comparison, 2024.
\newblock URL \url{https://docs.ultralytics.com/compare/yolov5-vs-yolov8/}.

\bibitem[nvi(2025{\natexlab{a}})]{nvidiamig}
Nvidia multi-instance gpu user guide, release r575, 2025{\natexlab{a}}.
\newblock URL \url{https://docs.nvidia.com/datacenter/tesla/mig-user-guide/}.

\bibitem[nvi(2025{\natexlab{b}})]{nvidiamps}
Nvidia multi-process service, release r575, 2025{\natexlab{b}}.
\newblock URL \url{https://docs.nvidia.com/deploy/mps/index.html}.

\bibitem[Ahmad et~al.(2024{\natexlab{a}})Ahmad, Guan, Friedman, Williams, Sitaraman, and Woo]{proteus}
Sohaib Ahmad, Hui Guan, Brian~D. Friedman, Thomas Williams, Ramesh~K. Sitaraman, and Thomas Woo.
\newblock Proteus: A high-throughput inference-serving system with accuracy scaling.
\newblock In \emph{Proceedings of the 29th ACM International Conference on Architectural Support for Programming Languages and Operating Systems, Volume 1}, ASPLOS '24, page 318–334, New York, NY, USA, 2024{\natexlab{a}}. Association for Computing Machinery.
\newblock ISBN 9798400703720.
\newblock \doi{10.1145/3617232.3624849}.
\newblock URL \url{https://doi.org/10.1145/3617232.3624849}.

\bibitem[Ahmad et~al.(2024{\natexlab{b}})Ahmad, Guan, and Sitaraman]{loki}
Sohaib Ahmad, Hui Guan, and Ramesh~K. Sitaraman.
\newblock Loki: A system for serving ml inference pipelines with hardware and accuracy scaling.
\newblock In \emph{Proceedings of the 33rd International Symposium on High-Performance Parallel and Distributed Computing}, HPDC '24, page 267–280, New York, NY, USA, 2024{\natexlab{b}}. Association for Computing Machinery.
\newblock ISBN 9798400704130.
\newblock \doi{10.1145/3625549.3658688}.
\newblock URL \url{https://doi.org/10.1145/3625549.3658688}.

\bibitem[Boerner et~al.(2023)Boerner, Deems, Furlani, Knuth, and Towns]{access}
Timothy~J. Boerner, Stephen Deems, Thomas~R. Furlani, Shelley~L. Knuth, and John Towns.
\newblock Access: Advancing innovation: Nsf’s advanced cyberinfrastructure coordination ecosystem: Services \& support.
\newblock In \emph{Practice and Experience in Advanced Research Computing 2023: Computing for the Common Good}, PEARC '23, page 173–176, New York, NY, USA, 2023. Association for Computing Machinery.
\newblock ISBN 9781450399852.
\newblock \doi{10.1145/3569951.3597559}.
\newblock URL \url{https://doi.org/10.1145/3569951.3597559}.

\bibitem[Chen et~al.(2025)Chen, Davis, Hanin, Bailis, Zaharia, Zou, and Stoica]{compoundaimodelselection}
Lingjiao Chen, Jared~Quincy Davis, Boris Hanin, Peter Bailis, Matei Zaharia, James Zou, and Ion Stoica.
\newblock Optimizing model selection for compound ai systems, 2025.
\newblock URL \url{https://arxiv.org/abs/2502.14815}.

\bibitem[Choi et~al.(2022)Choi, Lee, Kim, Park, Kwon, and Huh]{gpulet}
Seungbeom Choi, Sunho Lee, Yeonjae Kim, Jongse Park, Youngjin Kwon, and Jaehyuk Huh.
\newblock Serving heterogeneous machine learning models on {Multi-GPU} servers with {Spatio-Temporal} sharing.
\newblock In \emph{2022 USENIX Annual Technical Conference (USENIX ATC 22)}, pages 199--216, Carlsbad, CA, July 2022. USENIX Association.
\newblock ISBN 978-1-939133-29-53.
\newblock URL \url{https://www.usenix.org/conference/atc22/presentation/choi-seungbeom}.

\bibitem[{City of Bellevue}(2017)]{bellevuetrafficvideodataset}
{City of Bellevue}.
\newblock {Bellevue Traffic Video Dataset}.
\newblock {\url{https://github.com/City-of-Bellevue/TrafficVideoDataset}}, 2017.
\newblock Accessed: {August 4, 2025}.

\bibitem[Dhakal et~al.(2020)Dhakal, Kulkarni, and Ramakrishnan]{gslice}
Aditya Dhakal, Sameer~G Kulkarni, and K.~K. Ramakrishnan.
\newblock Gslice: controlled spatial sharing of gpus for a scalable inference platform.
\newblock In \emph{Proceedings of the 11th ACM Symposium on Cloud Computing}, SoCC '20, page 492–506, New York, NY, USA, 2020. Association for Computing Machinery.
\newblock ISBN 9781450381376.
\newblock \doi{10.1145/3419111.3421284}.
\newblock URL \url{https://doi.org/10.1145/3419111.3421284}.

\bibitem[Donnelly(2023)]{twitterfrankdonnelly}
Frank Donnelly.
\newblock Parsing the inter archive's twitter stream grab with python, 2023.
\newblock URL \url{https://atcoordinates.info/2023/04/30/parsing-the-internet-archives-twitter-stream-grab-with-python/}.

\bibitem[Ghafouri et~al.(2024)Ghafouri, Razavi, Salmani, Sanaee, Botran, Wang, Doyle, and Jamshidi]{ipa}
Saeid Ghafouri, Kamran Razavi, Mehran Salmani, Alireza Sanaee, Tania~Lorido Botran, Lin Wang, Joseph Doyle, and Pooyan Jamshidi.
\newblock [solution] ipa: Inference pipeline adaptation to achieve high accuracy and cost-efficiency.
\newblock \emph{Journal of Systems Research}, 4\penalty0 (1), April 2024.
\newblock ISSN 2770-5501.
\newblock \doi{10.5070/sr34163500}.
\newblock URL \url{http://dx.doi.org/10.5070/SR34163500}.

\bibitem[{Gurobi Optimization, LLC}(2024)]{gurobi}
{Gurobi Optimization, LLC}.
\newblock {Gurobi Optimizer Reference Manual}, 2024.
\newblock URL \url{https://www.gurobi.com}.

\bibitem[Hancock et~al.(2021)Hancock, Fischer, Lowe, Snapp-Childs, Pierce, Marru, Coulter, Vaughn, Beck, Merchant, Skidmore, and Jacobs]{jetstream2}
David~Y. Hancock, Jeremy Fischer, John~Michael Lowe, Winona Snapp-Childs, Marlon Pierce, Suresh Marru, J.~Eric Coulter, Matthew Vaughn, Brian Beck, Nirav Merchant, Edwin Skidmore, and Gwen Jacobs.
\newblock Jetstream2: Accelerating cloud computing via jetstream.
\newblock In \emph{Practice and Experience in Advanced Research Computing 2021: Evolution Across All Dimensions}, PEARC '21, New York, NY, USA, 2021. Association for Computing Machinery.
\newblock ISBN 9781450382922.
\newblock \doi{10.1145/3437359.3465565}.
\newblock URL \url{https://doi.org/10.1145/3437359.3465565}.

\bibitem[He et~al.(2015)He, Zhang, Ren, and Sun]{resnet}
Kaiming He, Xiangyu Zhang, Shaoqing Ren, and Jian Sun.
\newblock Deep residual learning for image recognition, 2015.
\newblock URL \url{https://arxiv.org/abs/1512.03385}.

\bibitem[{HPC Wire}(2019)]{inferencecostnvidia}
{HPC Wire}.
\newblock Aws to offer nvidia’s t4 gpus for ai inferencing, 2019.
\newblock URL \url{https://www.hpcwire.com/2019/03/19/aws-upgrades-its-gpu-backed-ai-inference-platform/}.

\bibitem[Hui et~al.(2024)Hui, Xu, Guo, and Shen]{esg}
Xinning Hui, Yuanchao Xu, Zhishan Guo, and Xipeng Shen.
\newblock Esg: Pipeline-conscious efficient scheduling of dnn workflows on serverless platforms with shareable gpus.
\newblock In \emph{Proceedings of the 33rd International Symposium on High-Performance Parallel and Distributed Computing}, HPDC '24, page 42–55, New York, NY, USA, 2024. Association for Computing Machinery.
\newblock ISBN 9798400704130.
\newblock \doi{10.1145/3625549.3658657}.
\newblock URL \url{https://doi.org/10.1145/3625549.3658657}.

\bibitem[Hui et~al.(2025)Hui, Xu, and Shen]{fluidfaas}
Xinning Hui, Yuanchao Xu, and Xipeng Shen.
\newblock Fluidfaas: A dynamic pipelined solution for serverless computing with strong isolation-based gpu sharing.
\newblock In \emph{The 34th International Symposium on High-Performance Parallel and Distributed Computing (HPDC ’25)}, Notre Dame, IN, USA, july 2025. ACM.
\newblock \doi{https://doi.org/10.1145/3731545.3731580}.
\newblock URL \url{https://research.csc.ncsu.edu/picture/publications/papers/hpdc2025.pdf}.

\bibitem[Kim et~al.(2020)Kim, Kim, Kong, and Yoon]{glowtts}
Jaehyeon Kim, Sungwon Kim, Jungil Kong, and Sungroh Yoon.
\newblock Glow-tts: A generative flow for text-to-speech via monotonic alignment search, 2020.
\newblock URL \url{https://arxiv.org/abs/2005.11129}.

\bibitem[Kim et~al.(2021)Kim, Kong, and Son]{vits}
Jaehyeon Kim, Jungil Kong, and Juhee Son.
\newblock Conditional variational autoencoder with adversarial learning for end-to-end text-to-speech, 2021.
\newblock URL \url{https://arxiv.org/abs/2106.06103}.

\bibitem[Kim et~al.(2024)Kim, Kwon, Song, Jo, Chen, Lai, and Chandra]{dream}
Seah Kim, Hyoukjun Kwon, Jinook Song, Jihyuck Jo, Yu-Hsin Chen, Liangzhen Lai, and Vikas Chandra.
\newblock Dream: A dynamic scheduler for dynamic real-time multi-model ml workloads.
\newblock In \emph{Proceedings of the 28th ACM International Conference on Architectural Support for Programming Languages and Operating Systems, Volume 4}, ASPLOS '23, page 73–86, New York, NY, USA, 2024. Association for Computing Machinery.
\newblock ISBN 9798400703942.
\newblock \doi{10.1145/3623278.3624753}.
\newblock URL \url{https://doi.org/10.1145/3623278.3624753}.

\bibitem[Kwon et~al.(2023)Kwon, Nair, Seo, Yik, Mohapatra, Zhan, Song, Capak, Zhang, Vajda, et~al.]{xrbench}
Hyoukjun Kwon, Krishnakumar Nair, Jamin Seo, Jason Yik, Debabrata Mohapatra, Dongyuan Zhan, Jinook Song, Peter Capak, Peizhao Zhang, Peter Vajda, et~al.
\newblock Xrbench: An extended reality (xr) machine learning benchmark suite for the metaverse.
\newblock \emph{Proceedings of Machine Learning and Systems}, 5:\penalty0 1--20, 2023.

\bibitem[Lee et~al.(2024)Lee, Seong, Kang, Lee, Na, Chun, Nikolopoulos, and Hong]{parvagpu}
Munkyu Lee, Sihoon Seong, Minki Kang, Jihyuk Lee, Gap-Joo Na, In-Geol Chun, Dimitrios Nikolopoulos, and Cheol-Ho Hong.
\newblock Parvagpu: Efficient spatial gpu sharing for large-scale dnn inference in cloud environments.
\newblock In \emph{SC24: International Conference for High Performance Computing, Networking, Storage and Analysis}, pages 1--14. IEEE, 2024.

\bibitem[Li et~al.(2022)Li, Patel, Samsi, Gadepally, and Tiwari]{misompsformig}
Baolin Li, Tirthak Patel, Siddharth Samsi, Vijay Gadepally, and Devesh Tiwari.
\newblock Miso: exploiting multi-instance gpu capability on multi-tenant gpu clusters.
\newblock In \emph{Proceedings of the 13th Symposium on Cloud Computing}, page 173–189. ACM, November 2022.
\newblock \doi{10.1145/3542929.3563510}.
\newblock URL \url{http://dx.doi.org/10.1145/3542929.3563510}.

\bibitem[Li et~al.(2023)Li, Samsi, Gadepally, and Tiwari]{clover}
Baolin Li, Siddharth Samsi, Vijay Gadepally, and Devesh Tiwari.
\newblock Clover: Toward sustainable ai with carbon-aware machine learning inference service.
\newblock In \emph{Proceedings of the International Conference for High Performance Computing, Networking, Storage and Analysis}, pages 1--15, 2023.

\bibitem[Lin et~al.(2015)Lin, Maire, Belongie, Bourdev, Girshick, Hays, Perona, Ramanan, Zitnick, and Dollár]{cocodataset}
Tsung-Yi Lin, Michael Maire, Serge Belongie, Lubomir Bourdev, Ross Girshick, James Hays, Pietro Perona, Deva Ramanan, C.~Lawrence Zitnick, and Piotr Dollár.
\newblock Microsoft coco: Common objects in context, 2015.
\newblock URL \url{https://arxiv.org/abs/1405.0312}.

\bibitem[Patterson et~al.(2022)Patterson, Gonzalez, Hölzle, Le, Liang, Munguia, Rothchild, So, Texier, and Dean]{energydatacentergoogle}
David Patterson, Joseph Gonzalez, Urs Hölzle, Quoc Le, Chen Liang, Lluis-Miquel Munguia, Daniel Rothchild, David So, Maud Texier, and Jeff Dean.
\newblock The carbon footprint of machine learning training will plateau, then shrink, 2022.
\newblock URL \url{https://arxiv.org/abs/2204.05149}.

\bibitem[Redmon et~al.(2016)Redmon, Divvala, Girshick, and Farhadi]{yolo}
Joseph Redmon, Santosh Divvala, Ross Girshick, and Ali Farhadi.
\newblock You only look once: Unified, real-time object detection, 2016.
\newblock URL \url{https://arxiv.org/abs/1506.02640}.

\bibitem[Romero et~al.(2021)Romero, Li, Yadwadkar, and Kozyrakis]{infaas}
Francisco Romero, Qian Li, Neeraja~J. Yadwadkar, and Christos Kozyrakis.
\newblock {INFaaS}: Automated model-less inference serving.
\newblock In \emph{2021 USENIX Annual Technical Conference (USENIX ATC 21)}, pages 397--411. USENIX Association, July 2021.
\newblock ISBN 978-1-939133-23-6.
\newblock URL \url{https://www.usenix.org/conference/atc21/presentation/romero}.

\bibitem[Shahrad et~al.(2020)Shahrad, Fonseca, Goiri, Chaudhry, Batum, Cooke, Laureano, Tresness, Russinovich, and Bianchini]{azureserverlessworkload}
Mohammad Shahrad, Rodrigo Fonseca, Inigo Goiri, Gohar Chaudhry, Paul Batum, Jason Cooke, Eduardo Laureano, Colby Tresness, Mark Russinovich, and Ricardo Bianchini.
\newblock Serverless in the wild: Characterizing and optimizing the serverless workload at a large cloud provider.
\newblock In \emph{2020 USENIX Annual Technical Conference (USENIX ATC 20)}, pages 205--218. USENIX Association, July 2020.
\newblock ISBN 978-1-939133-14-4.
\newblock URL \url{https://www.usenix.org/conference/atc20/presentation/shahrad}.

\bibitem[Shen et~al.(2019)Shen, Chen, Jin, Zhao, Kong, Philipose, Krishnamurthy, and Sundaram]{nexus}
Haichen Shen, Lequn Chen, Yuchen Jin, Liangyu Zhao, Bingyu Kong, Matthai Philipose, Arvind Krishnamurthy, and Ravi Sundaram.
\newblock Nexus: a gpu cluster engine for accelerating dnn-based video analysis.
\newblock In \emph{Proceedings of the 27th ACM Symposium on Operating Systems Principles}, SOSP '19, page 322–337, New York, NY, USA, 2019. Association for Computing Machinery.
\newblock ISBN 9781450368735.
\newblock \doi{10.1145/3341301.3359658}.
\newblock URL \url{https://doi.org/10.1145/3341301.3359658}.

\bibitem[Simonyan and Zisserman(2015)]{vgg}
Karen Simonyan and Andrew Zisserman.
\newblock Very deep convolutional networks for large-scale image recognition, 2015.
\newblock URL \url{https://arxiv.org/abs/1409.1556}.

\bibitem[Tan et~al.(2021)Tan, Li, Zhang, Cao, Qi, Liu, Zhu, and Guo]{migserving}
Cheng Tan, Zhichao Li, Jian Zhang, Yu~Cao, Sikai Qi, Zherui Liu, Yibo Zhu, and Chuanxiong Guo.
\newblock Serving dnn models with multi-instance gpus: A case of the reconfigurable machine scheduling problem, 2021.
\newblock URL \url{https://arxiv.org/abs/2109.11067}.

\bibitem[Tan and Le(2020)]{efficientnet}
Mingxing Tan and Quoc~V. Le.
\newblock Efficientnet: Rethinking model scaling for convolutional neural networks, 2020.
\newblock URL \url{https://arxiv.org/abs/1905.11946}.

\bibitem[Turkkan et~al.(2024)Turkkan, Murali, Harsha, Arora, Vanloo, and Narayanaswami]{ibmmigpacking}
Bekir Turkkan, Pavankumar Murali, Pavithra Harsha, Rohan Arora, Gerard Vanloo, and Chandra Narayanaswami.
\newblock Optimal workload placement on multi-instance gpus, 2024.
\newblock URL \url{https://arxiv.org/abs/2409.06646}.

\bibitem[Wang et~al.(2022)Wang, Yang, Hu, Li, Lin, Gan, Liu, Liu, and Wang]{microsoftgit}
Jianfeng Wang, Zhengyuan Yang, Xiaowei Hu, Linjie Li, Kevin Lin, Zhe Gan, Zicheng Liu, Ce~Liu, and Lijuan Wang.
\newblock Git: A generative image-to-text transformer for vision and language, 2022.
\newblock URL \url{https://arxiv.org/abs/2205.14100}.

\bibitem[Wu et~al.(2022)Wu, Raghavendra, Gupta, Acun, Ardalani, Maeng, Chang, Behram, Huang, Bai, Gschwind, Gupta, Ott, Melnikov, Candido, Brooks, Chauhan, Lee, Lee, Akyildiz, Balandat, Spisak, Jain, Rabbat, and Hazelwood]{metacarbonfootprint}
Carole-Jean Wu, Ramya Raghavendra, Udit Gupta, Bilge Acun, Newsha Ardalani, Kiwan Maeng, Gloria Chang, Fiona~Aga Behram, James Huang, Charles Bai, Michael Gschwind, Anurag Gupta, Myle Ott, Anastasia Melnikov, Salvatore Candido, David Brooks, Geeta Chauhan, Benjamin Lee, Hsien-Hsin~S. Lee, Bugra Akyildiz, Maximilian Balandat, Joe Spisak, Ravi Jain, Mike Rabbat, and Kim Hazelwood.
\newblock Sustainable ai: Environmental implications, challenges and opportunities, 2022.
\newblock URL \url{https://arxiv.org/abs/2111.00364}.

\bibitem[Zaharia et~al.(2024)Zaharia, Khattab, Chen, Davis, Miller, Potts, Zou, Carbin, Frankle, Rao, and Ghodsi]{compoundaisystems}
Matei Zaharia, Omar Khattab, Lingjiao Chen, Jared~Quincy Davis, Heather Miller, Chris Potts, James Zou, Michael Carbin, Jonathan Frankle, Naveen Rao, and Ali Ghodsi.
\newblock The shift from models to compound ai systems.
\newblock \url{https://bair.berkeley.edu/blog/2024/02/18/compound-ai-systems/}, 2024.

\bibitem[Zhang et~al.(2024)Zhang, Li, and Li]{miger}
Bowen Zhang, Shuxin Li, and Zhuozhao Li.
\newblock Miger: Integrating multi-instance gpu and multi-process service for deep learning clusters.
\newblock In \emph{Proceedings of the 53rd International Conference on Parallel Processing}, ICPP '24, page 504–513, New York, NY, USA, 2024. Association for Computing Machinery.
\newblock ISBN 9798400717932.
\newblock \doi{10.1145/3673038.3673089}.
\newblock URL \url{https://doi.org/10.1145/3673038.3673089}.

\bibitem[Zhang et~al.(2019)Zhang, Yu, Wang, and Yan]{markcloudinferenceserving}
Chengliang Zhang, Minchen Yu, Wei Wang, and Feng Yan.
\newblock {MArk}: Exploiting cloud services for {Cost-Effective}, {SLO-Aware} machine learning inference serving.
\newblock In \emph{2019 USENIX Annual Technical Conference (USENIX ATC 19)}, pages 1049--1062, Renton, WA, July 2019. USENIX Association.
\newblock ISBN 978-1-939133-03-8.
\newblock URL \url{https://www.usenix.org/conference/atc19/presentation/zhang-chengliang}.

\bibitem[Zhao et~al.(2025)Zhao, Hu, Yang, Gong, Shen, Zhao, Li, Liu, and Qu]{slopt}
Zhixin Zhao, Yitao Hu, Guotao Yang, Ziqi Gong, Chen Shen, Laiping Zhao, Wenxin Li, Xiulong Liu, and Wenyu Qu.
\newblock Slopt: Serving real-time inference pipeline with strict latency constraint.
\newblock \emph{IEEE Transactions on Computers}, 74\penalty0 (4):\penalty0 1431--1445, 2025.
\newblock \doi{10.1109/TC.2025.3528125}.

\bibitem[Zhou et~al.(2025)Zhou, Wan, Sun, Palangi, Iqbal, Vulić, Korhonen, and Arik]{multiagentsystem}
Han Zhou, Xingchen Wan, Ruoxi Sun, Hamid Palangi, Shariq Iqbal, Ivan Vulić, Anna Korhonen, and Sercan~O. Arik.
\newblock Multi-agent design: Optimizing agents with better prompts and topologies, 2025.
\newblock URL \url{https://arxiv.org/abs/2502.02533}.

\end{thebibliography}
\bibliographystyle{plainnat}


\newpage

\appendix

%



\section{Implementing Compound Inference Systems} \label{implementation}

To implement the compound inference systems for empirical evalutation, we extend ideas proposed by FluidFaaS \cite{fluidfaas} where each model instance performing a task runs as a separate CPU process, with the model instances satisfying data dependencies through Inter Process Communication (IPC) via shared memory on the host system. We run each model instance as separate CPU processes since the NVIDIA CUDA driver limits each CPU process to use only one MIG instance. 
Using the host shared memory for IPC instead of socket-based communication reduces communication overheads between the CPU processes incurred by (de)serialization, heavy OS system calls, and multiple memory copies by the kernel.

This implementation does not restrict \proj{}, and is just a design choice suitable for our testbed. The ideas of \proj{} can scale and be used for other production-ready implementations of compound inference systems where all components of the serving system and the model instances performing the tasks are containerized, hosted on different nodes or datacenters, and communicate with point-to-point socket-based protocols such as gRPC over the network.

\section{Task-graph-uninformed Baselines} \label{taskgraphuninformed}

For baselines that perform task-graph-uninformed resource budgeting, we need to statically divide the end-to-end latency SLO and the available resources among the tasks in a task-graph-uninformed manner. We use the multiplicative factors of the most accurate model variants to calculate the expected demand for each task when a serving system is task-graph-uninformed. We divide this expected demand by the maximum throughput that a model instance described by $(t,v,s,b)$ can achieve for the most accurate model variant performing task $t$, and them multiply it with the number of GPU slices in $s$. This gives the expected amount of resources required for task $t$. We assign task-wise resource budgets by dividing the total resources available in the ratio of the expected amount of resources required for each task.

For the task-wise latency SLO, we split the end-to-end latency SLO for each path in the task graph in the ratio of the highest inference latency that a model instance described by $(t,v,s,b)$ can incur for the most accurate model variant performing task $t$. For a task $t$ that belongs to multiple paths, the task-wise latency SLO is the minimum latency along all paths that $t$ is present.

Once these task-level latency SLO and resource budgets are allocated, each task is constrained to use the task-level limits. In contrast, task-graph-informed resource budgeting constrains the total end-to-end latency SLO and the total resources across all tasks in the compound inference system. The latency SLO and the resources can be flexibly apportioned among the constituent tasks in task-graph-informed resource budgeting.

We choose this approach to statically divide the end-to-end latency SLO and resources instead of naively dividing them equally among all the tasks to make strong baselines of the task-graph-uninformed configuration search spaces. A task-graph-uninformed serving system would not know the end-to-end SLO, but only the SLO for each task that a user would provide. The user is not aware of the MILP formulation, and has to statically determine the latency SLO for each task assuming the worst case latency that the serving system could choose. For the task-wise resource budget, we use this approach to find the ratio of resources for each task assuming that the serving system would choose to serve the demand with the configuration that can achieve the maximum throughput.





\end{document}